\documentclass[twocolumn,prl,lengthcheck,final,superscriptaddress,longbibliography]{revtex4-2}
\usepackage[english]{babel}
\usepackage[latin9]{inputenc}
\usepackage{color}
\usepackage{units}
\usepackage{amsmath}
\usepackage{amssymb}
\usepackage{graphicx}
\usepackage[unicode=true,pdfusetitle,
 bookmarks=true,bookmarksnumbered=false,bookmarksopen=false,
 breaklinks=false,pdfborder={0 0 1},backref=false,colorlinks=true]
 {hyperref}

\makeatletter

\usepackage{tikz}
\usetikzlibrary{calc}

\usepackage{graphicx}
\usepackage{dcolumn}
\usepackage{bm}
\usepackage{chngcntr}
\counterwithout{equation}{section}
\usepackage{units}
\usepackage{textcomp}
\usepackage{appendix}

\makeatother

\begin{document}
\title{Sensing Out-of-Equilibrium and Quantum Non-Gaussian environments via
induced Time-Reversal Symmetry Breaking on the quantum-probe dynamics}
\author{Martin Kuffer}
\affiliation{Centro At\'omico Bariloche, CONICET, CNEA, S. C. de Bariloche, 8400,
Argentina}
\affiliation{Instituto de Nanociencia y Nanotecnologia, CNEA, CONICET, S. C. de
Bariloche, 8400, Argentina}
\affiliation{Instituto Balseiro, CNEA, Universidad Nacional de Cuyo, S. C. de Bariloche,
8400, Argentina}
\author{Analia Zwick}
\affiliation{Centro At\'omico Bariloche, CONICET, CNEA, S. C. de Bariloche, 8400,
Argentina}
\affiliation{Instituto de Nanociencia y Nanotecnologia, CNEA, CONICET, S. C. de
Bariloche, 8400, Argentina}
\affiliation{Instituto Balseiro, CNEA, Universidad Nacional de Cuyo, S. C. de Bariloche,
8400, Argentina}
\author{Gonzalo A. \'Alvarez}
\email{gonzalo.alvarez@conicet.gov.ar}

\affiliation{Centro At\'omico Bariloche, CONICET, CNEA, S. C. de Bariloche, 8400,
Argentina}
\affiliation{Instituto de Nanociencia y Nanotecnologia, CNEA, CONICET, S. C. de
Bariloche, 8400, Argentina}
\affiliation{Instituto Balseiro, CNEA, Universidad Nacional de Cuyo, S. C. de Bariloche,
8400, Argentina}
\begin{abstract}
Advancing quantum sensing tools for investigating systems at atomic
and nanoscales is crucial for the progress of quantum technologies.
While numerous protocols employ quantum probes to extract information
from stationary or weakly coupled environments, the challenges intensify
at atomic- and nano-scales where the environment is inherently out-of-equilibrium
or strongly coupled with the sensor. We here prove that the time-reversal
symmetry in the quantum-sensor control dynamics is broken, when partial
information is probed from an environment that is out-of-equilibrium
with non-stationary fluctuations or is described by quantum non-Gaussian,
strongly coupled environmental correlations. We exploit this phenomenon
as a quantum sensing paradigm with proof-of-principle experimental
quantum simulations using solid-state nuclear magnetic resonance (NMR).
This introduces a signal contrast on a qubit-probe that quantifies
how far the sensed environment is from equilibrium or its quantum
non-Gaussian nature. Protocols are also presented to discern and filter
a variety of environmental properties including stationary, non-stationary
and non-Gaussian quantum noise fluctuations as a step toward sensing
the ubiquitous environments of a quantum-sensor at atomic and nanoscales.
\end{abstract}
\maketitle

\section{Introduction}

The progress on controlling single quantum systems at atomic and
nanometric scales has lead to the development of quantum technologies
\citep{Awschalom2018,Acin2018,Pelucchi2022}. Both the storage
and processing of quantum information in quantum devices suffer from
decoherence, the loss of quantum information as a function of time,
that distorts the encoded information \citep{Suter2016}. Nevertheless,
decoherence effects are a key information resource about the environment
that is exploited for designing novel quantum sensors with important
applications in geology, archaeology, material science, biology and
medicine \citep{Degen2017,Aslam2023,Zwick2023} .

These quantum probes have strong potential to enable measurement of
physical properties with unprecedented sensitivity, but more importantly
they allow probing spatial scales that are not accessible by classical
means, such as the atomic and nanoscales \citep{Aslam2023,Degen2017}.
They have already enabled the magnetometry of single neurons \citep{Barry2016}
and magnetic biomarkers with subcellular resolution \citep{Nie2021},
microscale and nanoscale detection of single molecules \citep{Lovchinsky2016,Schlipf2017},
and the probing of temperature-dependent biological processes in cells
and small organisms \citep{Neumann2013}. 

In the context of these scales, particularly when employing single
quantum sensors, environmental systems either manifest intrinsic out-of-equilibrium
features or can be unavoidable driven out-of-equilibrium via the quantum
feedback induced by the probe, both inducing non-stationary environmental
fluctuations \citep{Alvarez2015,Buca2019,Lewis-Swan2019,Landsman2019,Chalermpusitarak2020,Kuffer2022}.
Moreover, quantum probes at these scales can be strongly coupled to
its environment generating what is known as non-Gaussian effects \citep{Kotler2013,Norris2016,Sung2019,Wang2019,Wang2020a,Wang2021,Jerger2023}.
Conventional frameworks to describe the quantum open nature of these
sensors do not account for non-stationary environmental features and/or
non-Gaussian effects \citep{Alvarez2011,Bylander2011,Suter2016,Degen2017}.
Only recently, due to progress in quantum sensing technologies, frameworks
for sensing out-of-equilibrium and/or non-Gaussian environments with
quantum sensors have been introduced \citep{Norris2016,Sung2019,Chalermpusitarak2020,Wang2020a,Wang2021,Kuffer2022,Jerger2023}.

In this article, we delve into the realm of time reversal symmetry
to design quantum control sensing-paradigms of the ubiquitous environments
found at atomic and nanoscales. We demonstrate the time reversal symmetry
breaking in the quantum control of a qubit-sensor, specifically when
coupled to quantum non-Gaussian and/or out-of-equilibrium environmental
interactions. Leveraging this characteristic, arising from the partial
information observed by a quantum-probe, we present a novel quantum
sensing paradigm rooted in the design of time-asymmetric dynamical
control of the sensor. We name this technique SENSIT (Sensing of
Environmental Non-Symmetric Information due to  T-symmetry breaking)
and demonstrate it through experimental quantum simulations using
solid-state NMR.

In this context, we illustrate how the distance of the quantum environmental
state from equilibrium can be encoded onto a qubit-probe signal contrast.
Furthermore, we showcase the selective filtration of non-stationary
features with respect to stationary noise fluctuations. Additionally,
we introduce protocols that leverage this contrast to selectively
quantify quantum non-Gaussian features and non-equilibrium characteristics
of the environment. Overall, our work marks a step forward in the
practical application of quantum sensing technology, offering valuable
insights into the ubiquitous out-of-equilibrium and quantum non-Gaussian
environments encountered by quantum sensors.

\section{Decoding Environmental Information through Dynamically Controlled
Qubit-Sensor}

To demonstrate the quantum sensing paradigm based on the time reversal
symmetry breaking, we consider a dynamically controlled qubit-sensor
coupled to an environment that induces pure dephasing. This quantum
sensor platform is found in a variety of systems as in electrons in
diamonds \citep{Romach2015,SchmidLorch2015,Schlipf2017}, electronic
spins in nanoscale nuclear spin baths \citep{Ma2014}, quantum dots
\citep{Connors2022}, donors in silicon \citep{Muhonen2014}, superconducting
qubits \citep{Bylander2011}, trapped atoms \citep{Frey2017}, and
solid-, liquid- and gas-state NMR systems \citep{Alvarez2011,Zwick2023}.

\begin{figure*}[t]
\begin{centering}
\includegraphics[width=1\textwidth]{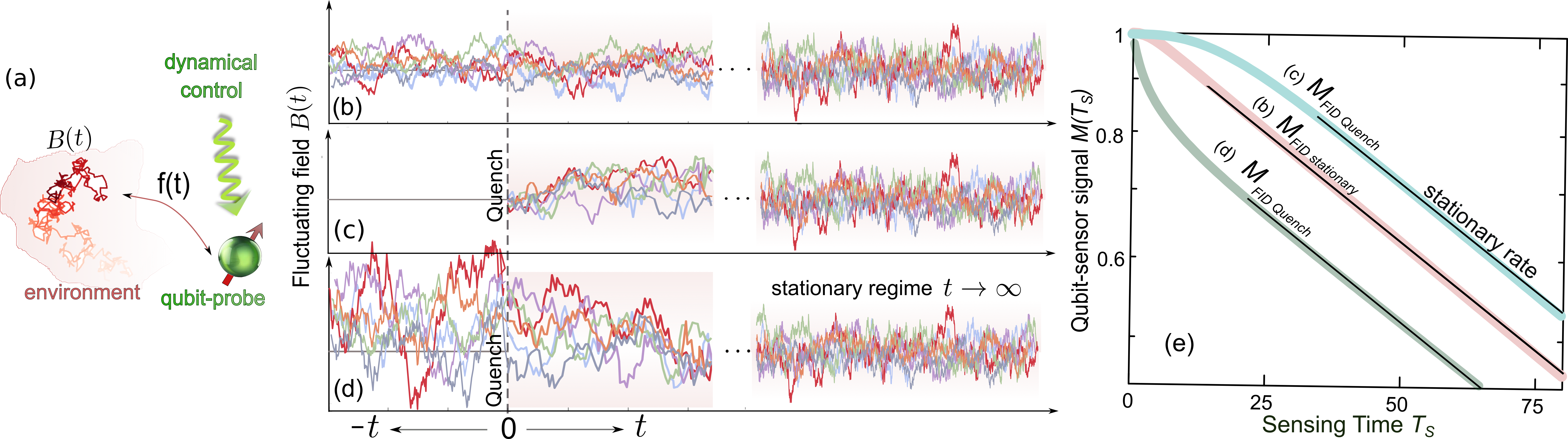}
\par\end{centering}
{\small{}\caption{\label{fig:Bquench}(a) Schematic illustration of the dynamically
modulated interaction $f(t)$, enabling the qubit-probe to sense its
environment. Realizations of a stochastic process of (b) a stationary
fluctuating field $B(t)$ as a function of time and (c,d) stochastic
processes representing an out-of-equilibrium environment due to a
quench by a change in the dynamics of the process at $t=0$. In (c),
prior to $t=0$, the fluctuating field remains fixed at $0$, evolving
stochastically for $t\protect\geq0$ and reaching a stationary regime
for $t\to\infty$. Notably, around $t\sim0$, the field exhibits a
smaller variance than at equilibrium, resulting in a comparatively
less impact of the qubit control on its signal. In (d) preceding $t=0$,
the fluctuating field has a greater variance than for $t\to\infty$,
evolving stochastically for $t\protect\geq0$ and attaining a stationary
regime for $t\to\infty$. Around $t\sim0$, the field has a greater
variance than at equilibrium, leading to a more pronounced impact
of the qubit control on its signal compared to the equilibrium state.
(e) Qubit-sensor signal $M(T_{s})$ as a function of the sensing time
$T_{s}$ for the fluctuating field displayed in panels (b-d). Distinctions
emerge at shorter times, yet the signals converge to the same decay
rate upon reaching the stationary regime. The decay-shift from the
stationary curve reflects the influence of an out-of-equilibrium environment
during the earlier times.}
}{\small\par}
\end{figure*}

Control on the qubit-sensor via dynamical decoupling effectively modulates
the strength of the qubit-environment interaction and can thus be
used to selectively encode environmental information on the signal
decay of the qubit-sensor \citep{Alvarez2011,Bylander2011}. In the
interaction picture with respect to the environmental evolution and
control of the qubit, the qubit-environment Hamiltonian is 
\begin{equation}
H_{SE}(t)=f(t)S_{z}B(t)\,,\label{eq:H_I(t)}
\end{equation}
where $S_{z}$ is the qubit-probe spin operator in the $z$ direction,
$B(t)$ is the noise operator representing the environmental fluctuating
degrees of freedom that induce dephasing on the qubit-probe, and
$f(t)$ is the qubit-environment interaction whose time dependence
is only due to the modulation induced by dynamical decoupling \citep{Viola1999a,Sagi2010,Malinowski2017a,Lange2010,Bylander2011,Alvarez2011}
(Fig. \ref{fig:Bquench}a). The fluctuating noise operator is given
by $B(t)=\mathrm{e}^{iH_{E}t}B\mathrm{e}^{-iH_{E}t}$, where $H_{E}$
is the environmental Hamiltonian and $B$ is the environmental degree
of freedom coupled to the qubit-probe.{\small{}}
{\small\par}

The evolution operator of the qubit-probe is thus $U(T_{s})=\mathcal{T}\mathrm{e}^{-i\int_{0}^{T_{s}}\mathrm{d}t\,f(t)S_{z}B(t)}$,
where $\mathcal{T}$ is the time-order superoperator and $T_{s}$
is the sensing time, the time during which the qubit-sensor dephases
due to sensing the environmental fluctuations. The probe observable
is the in-plane qubit-polarization that decays as $M(T_{s})=\frac{\left\langle S_{+}\otimes\mathbb{I}_{E}\,\rho(T_{s})\right\rangle }{\left\langle S_{+}\otimes\mathbb{I}_{E}\,\rho(0)\right\rangle }=\mathrm{e}^{-\mathcal{J}(T_{s})}$
due to the environment with the decoherence factor $\mathcal{J}$,
where $\rho(T_{s})$ is the density matrix of the full qubit-environment
system after the qubit has sensed the environment, $\left\langle \cdot\right\rangle =\text{tr}\left[\cdot\right]$,
and $S_{+}=S_{x}+iS_{y}$ the up spin operator.

We assume that, in the quantum sensing protocol, the qubit-environment
system is initially in a separable state when the probe is brought
into contact with the environment. Specifically, we consider the initial
state $\rho(0)=\rho_{0}=pS_{x}\otimes\rho_{E}$ with the qubit polarized
in the $x$ direction. We do not explicitly write terms proportional
to the qubit identity $\mathbb{I}_{S}$ in the density matrix, as
they do not contribute to the qubit-probe signal.

We perform a cumulant expansion of the decoherence factor $\mathcal{J}$
on the SE interaction coupling strength with the environment
\begin{equation}
\begin{array}{cc}
\mathcal{J}(T_{s})= & -\sum_{n}\frac{i^{n}}{n!}\int_{0}^{T_{s}}\mathrm{d}t_{1}\dots\int_{0}^{T_{s}}\mathrm{d}t_{n}\\
 & f(t_{1})\dots f(t_{n})W_{n}(t_{1},\dots,t_{n})
\end{array}\label{eq:J-de-Ws-maintext}
\end{equation}
where $W_{n}$ are the cumulants that completely characterize the
environment fluctuations felt by the qubit-sensor. They are defined
based on the environmental correlation functions
\begin{multline}
G_{n}(t_{1},\dots,t_{n})=\\
\frac{1}{2^{n-1}}\left\langle \left\{ B(t_{1}),\left\{ B(t_{2}),\left\{ \dots,B(t_{n})\right\} \dots\right\} \right\} \rho_{\text{E}}\right\rangle \label{eq:def-G-1}
\end{multline}
for $t_{1}\leq t_{2}\leq\dots\leq t_{n}$ with $\left\{ \cdot,\cdot\right\} $
the anti-commutator (see SI \ref{app:Signal-Decay}). 

\section{Environment-Induced Time-reversal Symmetry Breaking}

In quantum mechanics, the evolution operator $U(t)$ is unitary,
meaning it is invertible ($U^{\dagger}(t)=U^{-1}(t)$). When a quantum
system exhibits time-reversal symmetry, this is typically expressed
through the time reversal operator $T$, which is an anti-unitary
operator. Time reversal symmetry implies that $T$ commutes with the
system Hamiltonian $[T,H]=0$ and thus satisfies $T^{-1}U(t)T=U^{-1}(t)$
\citep{Domingos1979}.

At first glance, one might expect this symmetry to result in a corresponding
symmetry on the control operation of the qubit-probe dynamics. However,
we here demonstrate that when a dynamically controlled qubit-sensor
probes an environment, the partial information accessible to the sensor,
obtained through the partial trace of environmental degrees of freedom,
can unveil a breaking in the time-reversal symmetry of the control
function. In particular, we illustrate that the symmetry is disrupted,
when the noise operators of the environment, denoted as $B(t)$, fail
to commute at different times. This breakdown occurs explicitly when
the decoherence factor involves cumulants of order $n\geq3$. Moreover,
we also show that the symmetry is always broken if the environmental
fluctuations are non-stationary, indicative of non-equilibrium features
(Proof in SI \ref{app:T-symmetry}).

Quantum non-Gaussian noises thus induce time-reversal symmetry breaking
in the quantum control of the qubit-sensor. These noises are characteristic
of quantum environments strongly coupled to the sensor and operating
at low temperatures. The manifestation of symmetry breaking is absent
when the environment is weakly coupled validating the Gaussian approximation,
or at the high-temperature limit leading to the semiclassical field
approximation (see SI \ref{subapp:Time-reversal-environment}). Furthermore,
in the latter scenario, if the environmental fluctuations are non-Gaussian
but can be effectively described by classical fields, the time-reversal
symmetry remains protected.

A nonstationary environment does not have cumulants invariant under
time translation $\Delta t$, i.e. $W_{n}(t_{1},\dots,t_{n})\ne W_{n}(t_{1}+\Delta t,\dots,t_{n}+\Delta t)$.
In this case, for quantum or classical noise operators, the time-reversal
symmetry of the control is broken. Notice that time translation symmetry
is achieved when the environment reaches a stationary state, including
when it is in equilibrium. This is illustrated in Fig. \ref{fig:Bquench},
which compares stationary fluctuations of the field $B(t)$ (panel
b) with non-stationary fluctuations induced by quenches (panels c
and d). Time-reversal symmetry --in the control-- is broken in the
latter cases.

This asymmetry in time evolution is reflected in the dephasing of
the qubit-probe (Fig. \ref{fig:Bquench}e), where the dephasing near
the quench decays slower or faster compared to the stationary case,
depending on whether the noise fluctuation variance just before the
quench is lower or higher than in the stationary regime. This exemplifies
how one can exploit the manifestation of the time reversal symmetry
breaking of the environment on the qubit-probe dephasing.

The qubit-probe signal has time-reversal symmetry in the control
function if the cumulants in Eq. (\ref{eq:J-de-Ws-maintext}) satisfy
the symmetry condition $W_{n}(t_{1},\dots,t_{n})=(\pm1)^{n}W_{n}(T_{s}-t_{1},\dots,T_{s}-t_{n})$.
The sign $(\pm1)^{n}$ depends on the nature of the noise, i.e. when
the noise operator $B$ is an electric charge, electric field, gate
potential, etc. the sign corresponds to $1$, and in cases where it
is a magnetic field, magnetization, electric current, etc. corresponds
to $-1$. We thus call these cases of the electric- and magnetic-type,
respectively (details and proof in SI \ref{subapp:Time-reversal-environment}).
The cumulants exhibit this symmetry if and only if they possess time
translation symmetry, and the noise operators commute at different
times or their non-commutation is negligible.

Under these conditions, we prove that the qubit-probe signal satisfies
time-reversal symmetry in the control function
\begin{equation}
M_{f}=e^{-\mathcal{J}_{f}}=\begin{cases}
M_{f_{T}}=e^{-\mathcal{J}_{f_{T}}} & \text{electric-type}\\
M_{f_{T}}^{*}=e^{-\mathcal{J}_{f_{T}}^{*}} & \text{magnetic-type}
\end{cases}\,,\label{eq:TM-de-M}
\end{equation}
where $M_{f}$ is the signal measured when using the control function
$f$ and $M_{f_{T}}$ is that measured when using as control the
time reversal of $f$, i.e. $f_{T}(t)=f(T_{s}-t)$ (proof in SI \ref{subapp:Time-reversal-qubit}).
This demonstrates that the polarization of the qubit-probe remains
invariant, up to conjugation, under time reversal of the control sequence
determined by the control function $f$, when the cumulants exhibit
time-reversal symmetry. This time-reversal symmetry is a consequence
of \emph{stationary} correlation functions of an environment plus
negligible quantum noncommutativity effects guaranteed only by classical
fields, weak coupling, high temperature or Gaussianity. When either
of these conditions fails, the symmetry is broken, thus enabling the
measurement of quantum non-Gaussianity and/or non-stationary phenomena
due to out-of-equilibrium dynamics.

Therefore the argument $\Delta\mathcal{J}=\mathcal{J}_{f}-\mathcal{J}_{f_{T}}^{(*)}$
of the ratio between the corresponding qubit-probe signals $M_{f}/M_{f_{T}}^{(*)}=\exp\left\{ -\Delta\mathcal{J}\right\} $
is proportional to the degree of environmental time-reversal symmetry
breaking and/or the breaking of the time translation symmetry of the
environmental correlation functions (see SI \ref{app:SENSIT-Contrast}
and \ref{app:T-symmetry}). Here $^{(*)}$ corresponds to the complex
conjugation applied only for magnetic-type cases. The ratio $M_{f}/M_{f_{T}}^{(*)}$
is in general a complex number, where its modulus is given by the
even cumulants and the phase by the odd cumulants (see SI \ref{subapp:Time-reversal-qubit}).
Thus the modulus defines the \emph{SENSIT qubit-signal contrast,}
and the phase, the \emph{SENSIT qubit-phase contrast} (see SI \ref{app:SENSIT-Contrast})\emph{.
}In this article, our emphasis is on the SENSIT qubit-signal contrast
due to its greater accessibility and robustness in experiments. However,
in principle, either quantity can be employed.

The SENSIT qubit-signal contrast is\begin{widetext}

\begin{equation}
\text{Re}\Delta\mathcal{J}=-\sum_{n=0}^{\infty}\frac{\left(-1\right)^{n}}{\left(2n\right)!}\int_{0}^{T_{s}}\mathrm{d}t_{1}f(t_{1})\dots\int_{0}^{T_{s}}\mathrm{d}t_{2n}f(t_{2n})\,\Delta W_{2n}(t_{1},\dots,t_{2n})\,,\label{eq:dif_J}
\end{equation}
\end{widetext}where 
\begin{multline}
\Delta W_{2n}(t_{1},\dots,t_{2n})=\\
W_{2n}(t_{1},\dots,t_{2n})-W_{2n}(T_{s}-t_{1},\dots,T_{s}-t_{2n})\label{eq:def-Delta-W}
\end{multline}
is the difference between the $2n-$th cumulant at forward times $t_{i}$
and at the time reversed ones from the sensing time $T_{s}-t_{i}$.
These quantities serve as order parameters, gauging the extent of
time-reversal symmetry breaking in the qubit-control. The SENSIT contrast,
therefore, acts as a probe for these order parameters, with weights
determined by the control function $f(t)$. Notice that in Eq. (\ref{eq:dif_J}),
we performed a change in the time variables to convert the reversed
control $f_{T}(t)$ into $f(t)$, and this change the arrow of time
in the cumulants $W_{2n}(T_{s}-t_{1},\dots,T_{s}-t_{2n})$. 

The key to utilizing this quantification lies in its resilience against
any noise contribution to qubit-probe dephasing that does not induce
a time-reversal symmetry breaking, i.e. a contribution that is stationary
and such that the noise operators either commute at different times
or their non-commutation is negligible. This robustness is ensured
as the terms from $W_{2n}(t_{1},\dots,t_{2n})$ cancel out with those
in $W_{2n}(T_{s}-t_{1},\dots,T_{s}-t_{2n})$ and similarly for the
phase term (proof in SI \ref{subapp:Time-reversal-qubit}). This means
that if the sensed environment can be separated into two independent
parts, $a$ and $b$, such that only $a$ induces a time-reversal
symmetry breaking on the control, then the SENSIT contrast will exclusively
sense properties of $a$; while remaining completely independent of
$b$. Consequently, the SENSIT contrast exclusively responds to noise
sources that induce the time-reversal symmetry breaking of the qubit-probe
control (SI \ref{sec:Filter-out_stationary}). This may set an avenue
for pump and probe experiments, where a qubit that interacts with
a complex environment is used to sense just a part of it by first
driving the desired target subsystem out of equilibrium (pump) and
then use SENSIT to selectively detect just that subsystem (probe).

To give an example, the SENSIT contrast provides information about
the distance to equilibrium. In the simple but general case when the
environment state $\rho_{E}=\rho_{E}^{(0)}+\epsilon\rho_{E}^{(1)}$
is near to a stationary state $\rho_{E}^{(0)}$, with $\rho_{E}^{(1)}$
a constant perturbation, we demonstrate that $\Delta\mathcal{J\propto\epsilon}$
thus quantifying the distance to equilibrium (see SI \ref{sec:SENSIT-contrast-nearEquilibrium}).
In the particular cases illustrated in Fig. \ref{fig:Bquench}, the
SENSIT contrast $\text{Re}\Delta\mathcal{J}\propto\left(\sigma-\sigma_{0}\right)$,
where $\sigma$ the variance of the noise fluctuation before the quench,
and $\sigma_{0}$ the variance at the stationary regime achieved at
long times after the quench (see SI \ref{sec:SENSIT-Contrast-quenchedOU}).
This thus sets a paradigmatic example about how the SENSIT contrast
is proportional to the distance $\left(\sigma-\sigma_{0}\right)$
of the initial state of the environment to its stationary state at
equilibrium.

\section{Experimental detection of out-of-equilibrium states}

One of our main results is that the SENSIT contrast, being a readily
available quantity in experimental setups, serves as a sensing protocol
for detecting and characterizing out-of-equilibrium and/or quantum
non-Gaussian environments. To provide a concrete illustration of the
underlying principles of SENSIT and enhance the clarity of the introduced
general results, we experimentally performed solid-state NMR quantum
simulations on a Bruker Avance III HD 9.4T WB NMR spectrometer with
a $^{1}\text{H}$ resonance frequency of $400.15$ MHz and a $^{13}\text{C}$
resonance frequency of $100.61$ MHz. The $^{13}C$ nucleus plays
the role of the qubit-probe, and the surrounding $^{1}H$ nuclei are
considered the environment. Since $^{13}C$ is in natural abundance,
it is present in low concentration and all interaction between $^{13}C$
nuclei are negligible on the performed experiments (see SI \ref{submet:setup_experimental}
for details on the experimental setup). 

\begin{figure*}[t]
\includegraphics[width=1\textwidth]{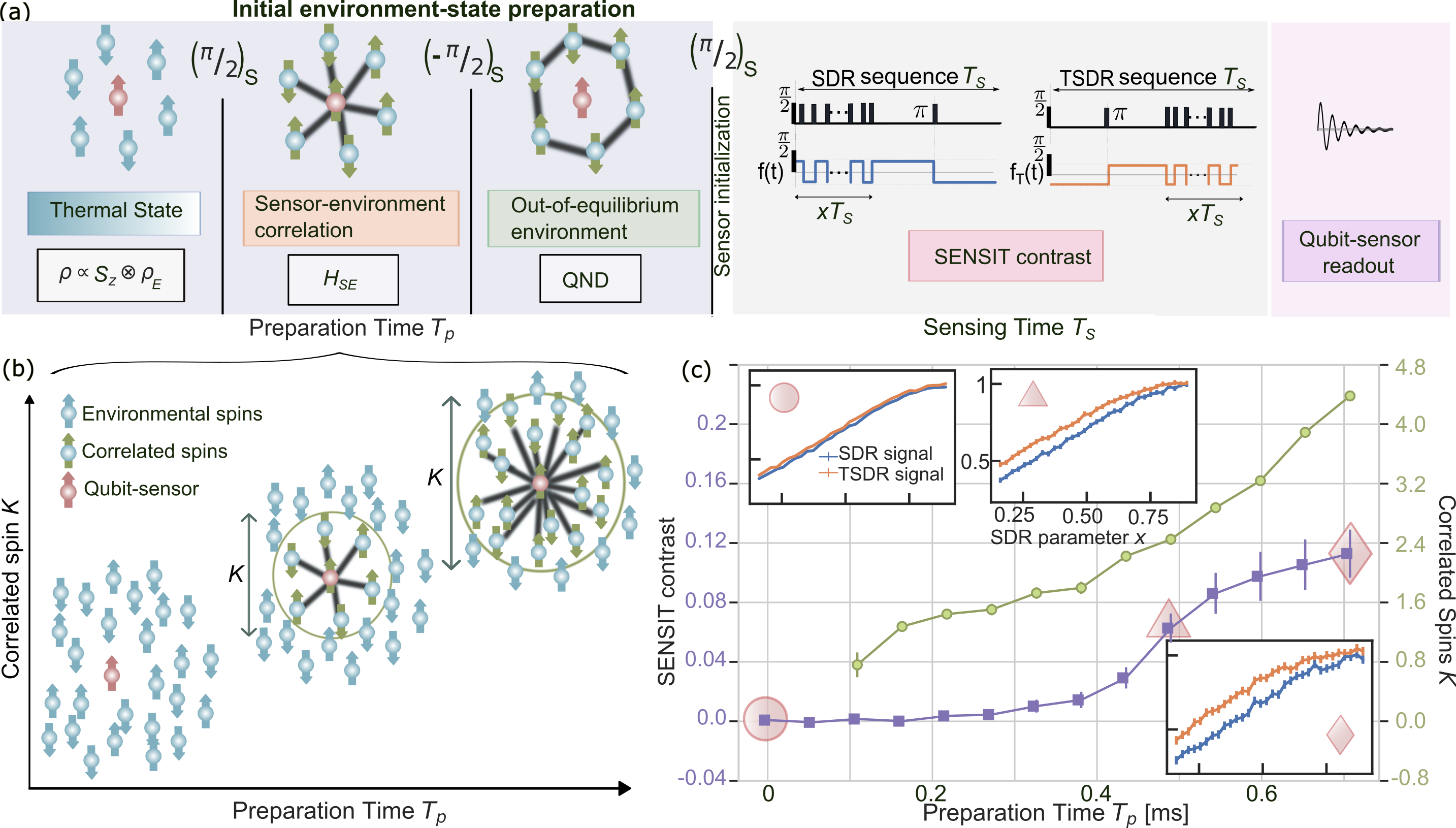}

\caption{Illustration of nonequilibrium sensing with SENSIT. (a) Experimental
protocol diagram implemented to induce nonequilibrium in the qubit
environment and its detection using the SENSIT technique. (b) Schematic
representation of the out-of-time-order commutator $K$, a well-established
measure quantifying the correlation of environmental spins with the
qubit during preparation. (c) Correlated spins $K$ and SENSIT contrast,
measured by the qubit, as a function of the preparation time $T_{p}$.
Error bars for $K$ are obtained from a fit, while those for the SENSIT
contrast account for the NMR detection sensitivity. The SENSIT contrast
$\int_{0}^{\frac{N-1}{N}}\mathrm{d}x\,\text{Re}\Delta\mathcal{J}(x)$
quantifies the environmental non-equilibrium degree by integrating
the difference of $\text{Re}\Delta\mathcal{J}(x)$ between SDR and
TSDR sequences over the parameter $x$. Insets show the signals measured
by SDR and TSDR control sequences for various preparation times (indicated
by geometrical symbols). Using $N=12$ pulses in our experiments,
the curves match at equilibrium ($T_{p}=0$), and diverge as the environment
deviates from equilibrium. This experimental demonstration highlights
the utilization of time-reversal symmetry breaking for measuring nonequilibrium
environments. }

\label{fig:Preparation_dependent}
\end{figure*}

The system is initially in thermal equilibrium, as represented in
the first step of the sensing protocol in Fig. \ref{fig:Preparation_dependent}a.
We induce an out-of-equilibrium state in the environment to generate
non-stationary noise fluctuations on the qubit-sensor. To achieve
this, we initially employ the qubit-probe to build up quantum correlations
between the qubit and the environmental spins during a preparation
time $T_{p}$, ensuring a localized spread of information near the
sensor (Fig. \ref{fig:Preparation_dependent}a,b). 
Then we performed a quantum nondemolition measurement on the qubit-probe
state mimicked by induced dephasing to erase the probe-environment
correlations \citep{Nakajima2019,Zwick2023}. This procedure leaves
the environment in a correlated out-of-equilibrium state (see third
step of Fig. \ref{fig:Preparation_dependent}a and SI \ref{submet:state-preparation}).
All these steps constitute the preparation of a nonequilibrium state
in the environment.

Subsequently, we initialize the qubit-probe state in a separable state
with the environment $\rho_{0}=pS_{x}\otimes\rho_{E}$ by the application
of a $\pi/2$ pulse on the qubit. After the initialization, we proceed
to implement non-equidistant dynamical decoupling sequences to manipulate
the time-reversal symmetry of the sequence. We observe the decay of
the qubit signal at the sensing time $T_{s}$, representing the duration
of the dynamical decoupling sequence, to quantify the SENSIT contrast.
This process is illustrated in the last two steps of Fig. \ref{fig:Preparation_dependent}a.

We employed the Selective Dynamical Recoupling (SDR) sequence \citep{Smith2012,Alvarez2013a},
chosen for its simplicity in constructing a time-asymmetric sequence
using only $\pi$-pulses, along with the flexibility of having a single
parameter that can be adjusted without changing $T_{s}$. This makes
SDR the most straightforward choice for our purposes. It consists
of a concatenation of a CPMG spin-echo train with $N-1$ rapid, spin-echo
train $\pi-$pulses between the  times $t=0$ and $t=xT_{s}$, and
a Hahn spin-echo sequence consisting of a single echo $\pi-$pulse
at the center between the times $xT_{s}$ and $T_{s}$. The SDR modulation
$f(t)$ is shown with the blue curve in Fig. \ref{fig:Preparation_dependent}a.
Here $x$ is a dimensionless parameter, that defines the asymmetry
of the SDR sequence interpolating between a single Hahn echo at $x=0$
and a CPMG sequence of $N$ equidistant pulses at $x=\frac{N-1}{N}$.
The time-reversed SDR (TSDR) sequence contains the inverse succession
in time of the $\pi-$pulses, consisting first of the single Hahn-echo
sequence between times $0$ and $(1-x)T_{s}$, followed by the $N-1$
CPMG pulses between the times $(1-x)T_{s}$ and $T_{s}$ (see the
orange curve for $f_{T}(t)$ in Fig. \ref{fig:Preparation_dependent}a
and SI \ref{submet:SDR}).

The insets in Fig. \ref{fig:Preparation_dependent}c illustrate the
qubit-probe signal following the SDR and TSDR modulations as a function
of $x$, for a fixed total sensing time and varying preparation
times \textbf{$T_{p}$ }for the initial out-of-equilibrium state in
the environment. The results demonstrate that the SDR and TSDR signals
are indistinguishable at $T_{p}=0$ when the environment is stationary
at equilibrium. Subsequently, they showcase how the signals progressively
increase their contrast as the environment is shifted further out
of equilibrium producing non-stationary noise fluctuations.

To get a single SENSIT-contrast quantification of the non-equilibrium
degree of the environmental state, we integrate the attenuation factors
of the qubit signals $\int_{0}^{\frac{N-1}{N}}\mathrm{d}x\,\text{Re}\Delta\mathcal{J}(x)$
over the parameter $x$ (main panel of Fig. \ref{fig:Preparation_dependent}c).
This contrast is proportional to the distance of the initial environmental
state from equilibrium, consequently increasing as a function of the
preparation time, with a greater number of environmental spins becoming
correlated (see SI \ref{app:T-symmetry}).

These experiments were performed at room temperature that represents
a high-temperature limit as the thermal energy is much larger than
the Zeeman energy of the spins (see SI \ref{submet:setup_experimental}),
therefore the time-reversal symmetry for quantum control described
in Eq. (\ref{eq:TM-de-M}) will only be broken if the environment
is out-of-equilibrium manifesting non-stationary noise fluctuations.
Thus here we demonstrate how the SENSIT contrast probes a quenched
state on the environment and determines how far from equilibrium it
is \citep{Alvarez2015,Wang2021,Jerger2023}.

\section{Comparing SENSIT Contrast and Out-of-Time-Order Correlation Metrics
for Out-of-Equilibrium States}

To have an alternative method for quantifying the out-of-equilibrium
degree, we employed a more established approach based on out-of-time-order
correlations (OTOC) measured with multiple quantum coherences \citep{Lewis-Swan2019,Xu2024}.
This method assesses the non-commutation degree of the evolved density
matrix during the preparation time in relation to the initial state
of the environment before preparation, as depicted in Fig. \ref{fig:Preparation_dependent}b.
This OTOC approach quantifies the effective number $K$ of environmental
spins that were correlated during the preparation step \citep{Alvarez2015,Dominguez2020}
(see SI \ref{submet:K} for details).

Figure \ref{fig:Preparation_dependent}c compares the number $K$
of correlated environmental spins to the qubit-probe during preparation,
and the SENSIT contrast measured from the qubit-probe. Both quantities
increase with the preparation time, manifesting a monotonous relation
between the established OTOC measure $K$ of non-equilibrium degree
and the proposed SENSIT contrast. Measuring the OTOC involves experimental
control over the environment, necessitating the ability to apply collective
rotations to environmental spins and reverse the many-body evolution
resulting from environmental spin-spin interaction (see SI \ref{submet:K}).
Consequently, while OTOC measurements are feasible in e.g. NMR experiments,
they are not readily available for most systems. In contrast, assessing
the SENSIT contrast only requires control over the qubit-probe, specifically
without the need for any control over the sensed environment.

\section{Probing quantum information scrambling with SENSIT contrast}

\begin{figure}[t]
\includegraphics[width=1\columnwidth]{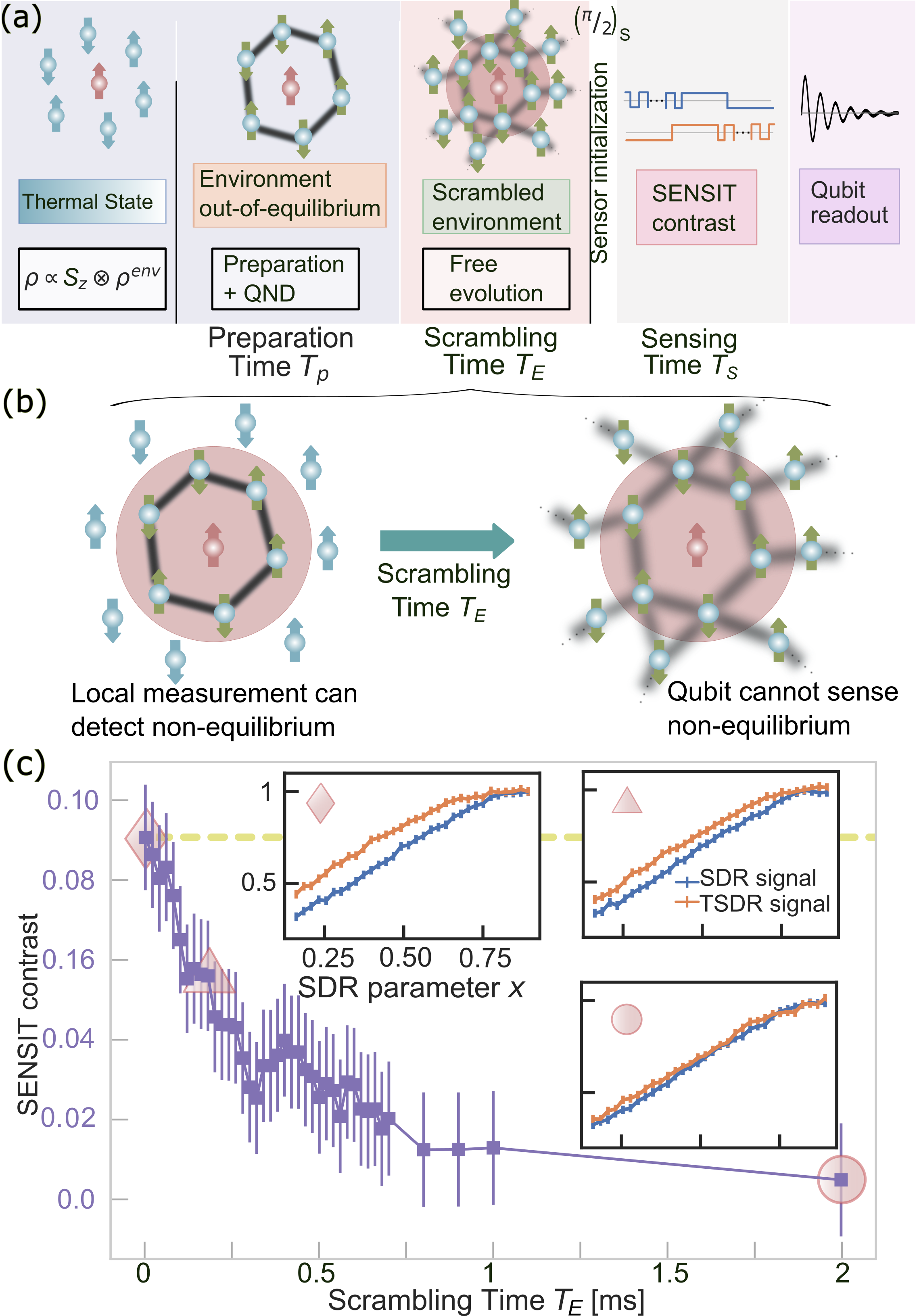}\caption{Sensing environmental scrambling with SENSIT. (a) Experimental protocol
diagram implemented to induce nonequilibrium in the qubit environment,
followed by scrambling and measurement using the SENSIT technique.
(b) Schematic representation of scrambling, the process encoding localized
information into non-local degrees of freedom, rendering it inaccessible
by local measurements. In the limit of large environmental scrambling
times $T_{E}$ , the qubit's ability to sense non-equilibrium conditions
diminishes. (c) Qubit-measured SENSIT contrast as a function of the
scrambling time $T_{E}$, enabling the distinction between non-stationary
environmental dynamics (experimental violet curve) and non-Gaussian
stationary noise (illustrative dashed yellow curve), which should
remain invariant under $T_{E}$. Error bars are due to the NMR detection
sensitivity. Insets display signals measured by the SDR and TSDR control
sequences for different scrambling times (marked by geometrical symbols).
$N=12$ pulses were used in our experiments. Note the separation of
curves at low scrambling times converging as the quench is scrambled
away.}

\label{fig:Scrambling_dependent}
\end{figure}

To illustrate how the SENSIT contrast, reliant on time-reversal symmetry
breaking, can selectively captures information about out-of-equilibrium
or quantum non-Gaussian environmental fluctuations over stationary
noise fluctuations, we measure the effect of environmental scrambling
\citep{Lewis-Swan2019,Xu2024}. In this context, we examine how the
quenched state information is erased due to scrambling, consequently
impacting the SENSIT contrast (Fig. \ref{fig:Scrambling_dependent}).

Quantum information scrambling is the encoding of an initial local
information into non-local degrees of freedom, in this case due to
the environmental dynamics driven by the Hamiltonian of the environment
$H_{E}$ (Fig. \ref{fig:Scrambling_dependent}b) \citep{Lewis-Swan2019,Xu2024}.
This in turn renders the information inaccessible by local measurements,
and it is related to the autothermalization\textbf{ }and quantum information
dynamics of the environment \citep{Alvarez2015,Li2017,Gaerttner2018,Landsman2019,Lewis-Swan2019,Niknam2020,Dominguez2020,Xu2024}.
Since the qubit-probe performs a local measurement, the measurements
after scrambling should match those found at equilibrium. The key
control source in this context involves introducing a waiting period,
denoted as the environmental scrambling time $T_{E}$, before implementing
the detection of the SENSIT contrast following the quenching of the
environment (Fig. \ref{fig:Scrambling_dependent}a). In the case
of a non-Gaussian environment that is stationary, the cumulant expansion
terms exhibit time translation symmetry, rendering the SENSIT contrast
invariant with respect to the waiting time. Conversely, an out-of-equilibrium
environment relaxes towards equilibrium as a function of the waiting---scrambling---time.

The decay of the SENSIT contrast, observed in Fig. \ref{fig:Scrambling_dependent}c,
is presented as a function of the environmental scrambling time $T_{E}$.
The insets provide the SDR and TSDR signals for three distinct scrambling
times. The contrast between the SDR and TSDR curves diminishes as
the quenched state is progressively scrambled away. These results
thus effectively showcase SENSIT's ability to selectively quantify
the extent to which the environment of the qubit-probe is departing
from equilibrium and evolving towards equilibrium. Importantly, SENSIT
enables the measurement of quantum information scrambling without
relying on environmental time reversions, a requirement typical need
in measurements based on OTOCs \citep{Lewis-Swan2019,Xu2024}.

\section{Summary and outlook}

While most quantum systems exhibit time reversal symmetry \citep{Domingos1979},
our research unveils a fundamental breakdown of time reversal symmetry
when a quantum sensor probes partial information from environments,
characterized by out-of-equilibrium, non-stationary dynamics or with
interactions containing quantum non-Gaussian correlations. This opens
a quantum sensing paradigm, offering a lens to explore ubiquitous
environments at the quantum level, where intrinsic out-of-equilibrium
dynamics prevail, driven either by inherent fluctuations or quantum
feedback induced by the probe \citep{Chalermpusitarak2020,Wang2021,Kuffer2022,Jerger2023}.
Moreover, it offers a tool for single sensors at atomic or nanoscales
that can be strongly coupled with the environment, thus generating
quantum non-Gaussian effects on the dephasing of the qubit-probe \citep{Jerger2023}.

Our findings gain practical significance in the realm of quantum sensing.
Specifically, within pump and probe schemes \citep{Li2024}, our
results open avenues for tailored measurements using qubit-probes,
enabling selective detection of the environmental degrees of freedom
being selectively pumped out of equilibrium. Moreover, our results
holds promise for characterizing noise of time crystals \citep{Greilich2024}
or Floquet systems \citep{Sridhar2024}, specially those with partially
broken time translation symmetry induced by external driving forces.
This offers insights into the dynamics of such systems, providing
a means to understand and quantify the intricate interplay between
the quantum probe and environments with complex, time-varying characteristics
\citep{Buca2019,Wang2021,Jerger2023}.

An additional strength of our work lies in its ability to capture
information scrambling of local operators, like the OTOCS, but notably
without requiring environmental time reversions \citep{Lewis-Swan2019,Xu2024}.
This unique feature positions our research as a valuable tool for
detecting persistent states characterized by long-lasting temporal
features, such as localized modes \citep{Shimasaki2024}. In essence,
our study not only contributes theoretical insights but also introduces
practical methodologies that can be harnessed to explore and manipulate
quantum systems in diverse and dynamic environments.

In conclusion, our study not only contributes fundamental insights
into the breakdown of control function time reversal symmetry in quantum
sensing but also presents a platform for practical applications. From
tailored measurements in complex samples to probing intriguing phenomena
in time crystal and Floquet systems, the potential impact of our findings
offers an alternative avenue in the realm of quantum sensing technologies.
This work paves the way for future research directions and applications,
emphasizing the dynamic interplay between quantum sensors and complex,
out-of-equilibrium environments.

\section*{ACKNOWLEDGMENTS}

This work was supported by CNEA; CONICET; ANPCyT-FONCyT PICT-2017-3156,
PICT-2017-3699, PICT-2018-4333, PICT-2021-GRF-TI-00134, PICT-2021-I-A-00070;
PIP-CONICET (11220170100486CO); UNCUYO SIIP Tipo I 2019-C028, 2022-C002,
2022-C030; Instituto Balseiro; Collaboration programs between the
MINCyT (Argentina) and, MAECI (Italy) and MOST (Israel).
\renewcommand\appendixpagename{Appendix}

\newpage{}

\appendix
\begin{widetext}

\section*{}

\setcounter{equation}{0}

\counterwithout{equation}{section}
\renewcommand{\theequation}{A.\arabic{equation}}

\section{Cumulant expansion for the qubit-probe signal decay\label{app:Signal-Decay}}

In this section we derive the cumulant expansion of the decoherence
factor $\mathcal{J}$ given by Eq. (\ref{eq:J-de-Ws-maintext}) in
the main text. Initially, we describe the signal decay of a controlled
qubit-probe that dephases due to the influence of an arbitrary quantum
environment. We then express it as a function of the correlation functions
of the environmental fluctuation, subsequently deriving the cumulant
expansion representation.

\subsection{Qubit-probe dephasing in terms of the environmental correlation functions}

We denote the initial state of the full system as $\rho(t=0)=\rho_{0}$
and we assume it is separable 
\begin{equation}
\rho_{0}=pS_{x}\otimes\rho_{E},\label{eq:rho_inicial}
\end{equation}
where $p$ is the polarization of the qubit and the qubit-probe state
$S_{x}$ is in the $x$ direction. Notice that, we do not explicitly
write terms proportional to the qubit identity $\mathbb{I}_{S}$ in
the density matrix, as they do not contribute to the qubit-probe signal.
We calculate the decay of in-plane polarization $M=\frac{2}{p}\left\langle S_{+}\otimes\mathbb{I}_{E}U(T_{s})\rho_{0}U^{\dagger}(T_{s})\right\rangle $
of the qubit sensor after dephasing 
\begin{equation}
M=\frac{2}{p}\left\langle S_{+}\otimes\mathbb{I}_{E}U(T_{s})\rho_{0}U^{\dagger}(T_{s})\right\rangle =\frac{2}{p}\left\langle S_{+}\otimes\mathbb{I}_{E}U(T_{s})pS_{x}\otimes\rho_{E}U^{\dagger}(T_{s})\right\rangle \label{eq:Mtilde_general}
\end{equation}
where $\mathbb{I}_{E}$ is the environmental identity matrix and
$U(T_{s})$ is the evolution operator. As described in the maintext,
$U(T_{s})=\mathcal{T}\mathrm{e}^{-i\int_{0}^{T_{s}}\mathrm{d}t\,f(t)S_{z}B(t)}$,
and evaluating the trace\textbf{ }over the qubit-degrees of freedom
in Eq. (\ref{eq:Mtilde_general})\textbf{,} we find 
\begin{equation}
\begin{array}{cc}
M & =\left\langle \left(\mathcal{T}\mathrm{e}^{\frac{i}{2}\int_{0}^{T_{s}}\mathrm{d}t\,f(t)B(t)}\right)\rho_{E}\left(\mathcal{T}\mathrm{e}^{-\frac{i}{2}\int_{0}^{T_{s}}\mathrm{d}t\,f(t)B(t)}\right)^{\dagger}\right\rangle \\
 & =\left\langle \left(\mathcal{T}\mathrm{e}^{\frac{i}{2}\int_{0}^{T_{s}}\mathrm{d}t\,f(t)B(t)}\right)\rho_{E}\left(\tilde{\mathcal{T}}\mathrm{e}^{\frac{i}{2}\int_{0}^{T_{s}}\mathrm{d}t\,f(t)B(t)}\right)\right\rangle \,,
\end{array}\label{eq:Generador_Gs}
\end{equation}
where $\tilde{\mathcal{T}}$ is the anti-time-order superoperator.
Considering $M$ as a functional of $f$, we can perform a functional
Taylor expansion in $f$ to obtain
\begin{equation}
M=\sum_{n}\frac{1}{n!}\int\mathrm{d}t_{1}\dots\int\mathrm{d}t_{n}\,f(t_{1})\dots f(t_{n})\left.\left[\frac{\delta}{\delta f(t_{1})}\dots\frac{\delta}{\delta f(t_{n})}M\right]\right|_{f=0}\,,
\end{equation}
where $\frac{\delta}{\delta f(t)}$ represents functional differentiation
with respect to $f(t)$. Defining the environmental correlation functions
as 
\begin{align}
G_{n}(t_{1},\dots,t_{n})= & \left.\left(-i\right)^{n}\left[\frac{\delta}{\delta f(t_{1})}\dots\frac{\delta}{\delta f(t_{n})}M\right]\right|_{f=0}\\
= & \left.\left(-i\right)^{n}\left[\frac{\delta}{\delta f(t_{1})}\dots\frac{\delta}{\delta f(t_{n})}\left\langle \left(\mathcal{T}\mathrm{e}^{\frac{i}{2}\int_{0}^{T_{s}}\mathrm{d}t\,f(t)B(t)}\right)\rho_{E}\left(\mathcal{T}\mathrm{e}^{-\frac{i}{2}\int_{0}^{T_{s}}\mathrm{d}t\,f(t)B(t)}\right)^{\dagger}\right\rangle \right]\right|_{f=0},
\end{align}
 we find that 

\begin{equation}
M=\sum_{n}\frac{i^{n}}{n!}\int_{0}^{T_{s}}\mathrm{d}t_{1}\dots\mathrm{d}t_{n}\,f(t_{1})\dots f(t_{n})G_{n}(t_{1},\dots,t_{n})\,.\label{eq:M-de-G}
\end{equation}
Notice that the correlation functions $G_{n}(t_{1},\dots,t_{n})$
are symmetric under exchange of their arguments, i.e. $G_{n}(t_{1},\dots t_{i},\dots,t_{j},\dots,t_{n})=G_{n}(t_{1},\dots t_{j},\dots,t_{i},\dots,t_{n})$.
They depend solely on the environmental time-dependent noise operator
$B(t)$ and initial state $\rho_{E}$ as
\begin{equation}
G_{n}(t_{1},\dots,t_{n})=\frac{1}{2^{n-1}}\left\langle \left\{ B(t_{1}),\left\{ B(t_{2}),\left\{ \dots,B(t_{n})\right\} \dots\right\} \right\} \rho_{E}\right\rangle ,\label{eq:def-G}
\end{equation}
for $t_{1}\leq t_{2}\leq\dots\leq t_{n}$. These correlation functions
are the mean value of the nested anti-commutators $\left\{ \cdot,\cdot\right\} $
of the noise operator in the environmental state $\rho_{E}$. The
lowest order ones are 
\begin{align*}
G_{0}= & 1\,,\\
G_{1}(t_{1})= & \left\langle B(t_{1})\rho_{E}\right\rangle \,,\\
G_{2}(t_{1},t_{2})= & \frac{1}{2}\left\langle \left\{ B(t_{1}),B(t_{2})\right\} \rho_{E}\right\rangle \,,\\
G_{3}(t_{1},t_{2},t_{3})= & \frac{1}{4}\left\langle \left\{ B(t_{1}),\left\{ B(t_{2}),B(t_{3})\right\} \right\} \rho_{E}\right\rangle \,,
\end{align*}
for $t_{1}\leq t_{2}\leq t_{3}$. While this procedure is perturbative,
the knowledge of the correlation functions completely determines the
decay of the qubit even in the non-perturbative regime. 

\subsection{The cumulant expansion for the qubit-signal decay\label{subsec:The-cumulant-expansion}}

The correlation functions $G_{n}$ completely describe the decay of
the qubit-probe. However, certain effects can be more easily described
using the cumulants, e.g. non-Gaussianity, separability of the effects
of independent processes, exponential signal decay and quantification
of the control time-reversal symmetry breaking. Expanding in a perturbation
series the decoherence factor $\mathcal{J}$ with $M=\mathrm{e}^{-\mathcal{J}}$,
 we obtain Eq. (\ref{eq:J-de-Ws-maintext}) of the main text
\begin{equation}
\mathcal{J}=-\sum_{n}\frac{i^{n}}{n!}\int_{0}^{T_{s}}\mathrm{d}t_{1}\dots\int_{0}^{T_{s}}\mathrm{d}t_{n}\,f(t_{1})\dots f(t_{n})W_{n}(t_{1},\dots,t_{n})\,,\label{eq:J-de-W}
\end{equation}
where $W_{n}$ are the cumulants of the environmental correlation
functions. Then by expanding $M=\sum_{n}\frac{(-1)^{n}}{n!}\mathcal{J}^{n}$,
we find the connection between the cumulants $W_{n}$ and the correlation
functions $G_{n}$
\begin{align*}
W_{0}= & 0\\
W_{1}(t_{1})= & G_{1}(t_{1})\\
W_{2}(t_{1},t_{2})=G_{2}(t_{1} & ,t_{2})-W_{1}(t_{1})W_{1}(t_{2})\\
W_{3}(t_{1},t_{2},t_{3})=G_{3}(t_{1},t_{2},t_{3})-W_{2}(t_{1},t_{2})W_{1}(t_{3})- & W_{1}(t_{1})W_{2}(t_{2},t_{3})-W_{1}(t_{2})W_{2}(t_{3},t_{1})-W_{1}(t_{1})W_{1}(t_{2})W_{1}(t_{3})\,.
\end{align*}

The general guideline for constructing the $n$-th order cumulant
$W_{n}$, involves subtracting from $G_{n}$ all possible combinations
of functions generated by taking products of cumulants with orders
less than $n$ \citep{vanKampen1992}. Specifically, for a Gaussian
environment, Wick's theorem dictates that $W_{n}=0$ for $n>2$ \citep{vanKampen1992}.

Since the correlation functions $G_{n}$ are real, the cumulants $W_{n}$
are also real. Therefore the cumulants of odd order contribute to
the qubit signal phase $\text{arg}M=-\text{Im}\left(\mathcal{J}\right)$
based on Eq. (\ref{eq:J-de-W})
\begin{equation}
\text{Im}\left(\mathcal{J}\right)=-\sum_{n}\frac{\left(-1\right)^{n}}{\left(2n+1\right)!}\int_{0}^{T_{s}}\mathrm{d}t_{1}\dots\int_{0}^{T_{s}}\mathrm{d}t_{2n+1}\,f(t_{1})\dots f(t_{2n+1})W_{2n+1}(t_{1},\dots,t_{2n+1}),\label{eq:imJ-de-W}
\end{equation}
and the cumulants of even order to the absolute value of the qubit-signal
$\left|M\right|=\mathrm{e}^{-\text{Re}\left(\mathcal{J}\right)}$,
with
\begin{equation}
\text{Re}\left(\mathcal{J}\right)=-\sum_{n}\frac{\left(-1\right)^{n}}{\left(2n\right)!}\int_{0}^{T_{s}}\mathrm{d}t_{1}\dots\int_{0}^{T_{s}}\mathrm{d}t_{2n}\,f(t_{1})\dots f(t_{2n})W_{2n}(t_{1},\dots,t_{2n})\,.\label{eq:reJ-de-W}
\end{equation}

\section{SENSIT Contrast\label{app:SENSIT-Contrast}}

We here calculate the SENSIT contrast based on the comparison between
the qubit-probe signal $M_{f}$, when the control modulating function
is $f(t)$, with the signal $M_{f_{T}}$, for the time reversed control
$f_{T}(t)=f(T_{s}-t)$. The corresponding decoherence factors based
on Eq. (\ref{eq:J-de-W}) are
\begin{align*}
\mathcal{J}_{f}= & -\sum_{n}\frac{i^{n}}{n!}\int_{0}^{T_{s}}\mathrm{d}t_{1}\dots\int_{0}^{T_{s}}\mathrm{d}t_{n}\,f(t_{1})\dots f(t_{n})W_{n}(t_{1},\dots,t_{n})\,,\\
\mathcal{J}_{f_{T}}= & -\sum_{n}\frac{i^{n}}{n!}\int_{0}^{T_{s}}\mathrm{d}t_{1}\dots\int_{0}^{T_{s}}\mathrm{d}t_{n}\,f_{T}(t_{1})\dots f_{T}(t_{n})W_{n}(t_{1},\dots,t_{n})\,,
\end{align*}
respectively. Changing variables we can rewrite the last equation
as $\mathcal{J}_{f_{T}}=-\sum_{n}\frac{\left(-i\right)^{n}}{n!}\int_{0}^{T_{s}}\mathrm{d}t_{1}\dots\int_{0}^{T_{s}}\mathrm{d}t_{n}\,f(t_{1})\dots f(t_{n})W_{n}(T_{s}-t_{1},\dots,T_{s}-t_{n})$.

Then the difference  between the attenuation factors is $\Delta\mathcal{J}=\mathcal{J}_{f}-\mathcal{J}_{f_{T}}^{(*)}$,
where $^{(*)}$ is complex conjugation applied only for magnetic-type
cases. The distinction between electric- and magnetic-type cases is
described in SI \ref{app:T-symmetry}. In terms of the cumulant differences,
we obtain 
\begin{equation}
\Delta\mathcal{J}=-\sum_{n}\frac{i^{n}}{\left(n\right)!}\int_{0}^{T_{s}}\mathrm{d}t_{1}\dots\int_{0}^{T_{s}}\mathrm{d}t_{n}\,f(t_{1})\dots f(t_{n})\,\Delta W_{n}(t_{1},\dots,t_{n})\,,\label{eq:SENSIT-completo-de-DW}
\end{equation}
where the cumulant differences are
\[
\Delta W_{n}(t_{1},\dots,t_{n})=W_{n}(t_{1},\dots,t_{n})-\left(\pm1\right)^{n}W_{n}(T_{s}-t_{1},\dots,T_{s}-t_{n})\,,
\]
with $\pm1$ being $1$ for electric- and $-1$ for magnetic-type
cases. For cumulants of even order, this distinction is irrelevant
as $\left(\pm1\right)^{2n}=1$, leading to Eq. (\ref{eq:def-Delta-W})
in the main text.

As stated in the main text, they measure the degree of time reversal
symmetry breaking of the control operations of the qubit-probe, acting
as a kind of order parameters. As shown in SI \ref{subsec:The-cumulant-expansion},
$(-i)^{n}W_{n}$ is real for $n$ and imaginary for $n$ odd. Consequently,
$\Delta\mathcal{J}$ separates into real and imaginary components.
The real (resp. imaginary) component is influenced only by cumulants
of even (resp. odd) order, corresponding to a difference in the magnitude
(resp. phase) of the qubit-probe signals $M_{f}$ and $M_{f_{T}}$.
Explicitly real component
\[
\text{Re}\Delta\mathcal{J}=\ln\left|\frac{M_{f_{T}}}{M_{f}^{(*)}}\right|=\ln\left|\frac{M_{f_{T}}}{M_{f}}\right|=-\sum_{n}\frac{\left(-1\right)^{n}}{\left(2n\right)!}\int_{0}^{T_{s}}\mathrm{d}t_{1}\dots\int_{0}^{T_{s}}\mathrm{d}t_{2n}\,f(t_{1})\dots f(t_{2n})\Delta W_{2n}(t_{1},\dots,t_{2n})\,
\]
corresponds to a difference in the magnitude of the qubit-probe signals,
and it is defined by cumulant differences of even order $\Delta W_{2n}$.
This thus demonstrates Eq. (\ref{eq:dif_J}) of the main text. Analogously,
the imaginary component
\[
\text{Im}\Delta\mathcal{J}=\text{arg}\left(\frac{M_{f_{T}}}{M_{f}^{(*)}}\right)=-\sum_{n=0}^{\infty}\frac{\left(-1\right)^{n}}{\left(2n+1\right)!}\int_{0}^{T_{s}}\mathrm{d}t_{1}f(t_{1})\dots\int_{0}^{T_{s}}\mathrm{d}t_{2n+1}f(t_{2n+1})\Delta W_{2n+1}(t_{1},\dots,t_{2n})
\]
corresponds to a difference in the phases of the qubit signals, and
it is defined by the cumulant differences of odd order $\Delta W_{2n+1}$.

\section{Conditions in the environment for attaining time-reversal symmetry
of the qubit-probe control \label{app:T-symmetry}}

In this Section, we analyze the conditions required on the environment
to satisfy the time-reversal symmetry of the control on the qubit-probe.
Based on the expression for the SENSIT contrast of Eq. (\ref{eq:SENSIT-completo-de-DW})
and Eq. (\ref{eq:dif_J}) of the maintext, the cumulant difference
$\Delta W_{n}(t_{1},\dots,t_{2n})=0$ should vanish.

To obtain $\Delta W_{n}(t_{1},\dots,t_{n})=0$ for all sensing times
$T_{s}$, it is required for the cumulant functions to have time translation
symmetry and that $W_{n}(t_{1},\dots,t_{2n})=\left(\pm1\right)^{n}W_{n}(-t_{1},\dots,-t_{2n})$,
where the sign $\left(\pm1\right)^{n}$ depends on the nature of the
noise operator and will be explained later in this section. The first
condition is attained if the environmental noise operator fluctuations
are stationary. The second one if they have time-reversal symmetry.
In Sec. \ref{subapp:Time-shift-environment} we thus evaluate the
effects of the time translation symmetry of stationary environments
and in Sec. \ref{subapp:Time-reversal-environment} we determine the
effects of time-reversal symmetry in the environment.  \emph{}

\subsection{Time translation symmetry of the correlation functions and cumulants:
a stationary environment \label{subapp:Time-shift-environment}}

To obtain $\Delta W_{n}(t_{1},\dots,t_{n})=0$, we need the cumulants
to have time translation symmetry
\[
W_{n}(t_{1}+\Delta t,\dots,t_{n}+\Delta t)=W_{n}(t_{1},\dots,t_{n})\,.
\]
As the cumulants are written in terms of the environment correlation
functions $G_{n}$ of Eq. (\ref{eq:def-G}), we thus need them to
have time translation symmetry $G_{n}(t_{1}+\Delta t,\dots,t_{n}+\Delta t)=G_{n}(t_{1},\dots,t_{n})$.
As the noise operators are defined by $B(t)=\mathrm{e}^{itH_{E}}B\mathrm{e}^{-itH_{E}}$,
where $H_{E}$ is the environment Hamiltonian, we obtain
\begin{align*}
G_{n}(t_{1}+\Delta t,\dots,t_{n}+ & \Delta t)=\frac{1}{2^{n-1}}\left\langle \left\{ B(t_{1}+\Delta t),\left\{ B(t_{2}+\Delta t),\left\{ \dots,B(t_{n}+\Delta t)\right\} \dots\right\} \right\} \rho_{E}\right\rangle \\
=\frac{1}{2^{n-1}} & \left\langle \left\{ \mathrm{e}^{i\Delta tH_{E}}B(t_{1})\mathrm{e}^{-i\Delta tH_{E}},\left\{ \mathrm{e}^{i\Delta tH_{E}}B(t_{2})\mathrm{e}^{-i\Delta tH_{E}},\left\{ \dots,\mathrm{e}^{i\Delta tH_{E}}B(t_{n})\mathrm{e}^{-i\Delta tH_{E}}\right\} \dots\right\} \right\} \rho_{E}\right\rangle \\
=\frac{1}{2^{n-1}} & \left\langle \mathrm{e}^{i\Delta tH_{E}}\left\{ B(t_{1}),\left\{ B(t_{2}),\left\{ \dots,B(t_{n})\right\} \dots\right\} \right\} \mathrm{e}^{-i\Delta tH_{E}}\rho_{E}\right\rangle \\
=\frac{1}{2^{n-1}} & \left\langle \left\{ B(t_{1}),\left\{ B(t_{2}),\left\{ \dots,B(t_{n})\right\} \dots\right\} \right\} \mathrm{e}^{-i\Delta tH_{E}}\rho_{E}\mathrm{e}^{i\Delta tH_{E}}\right\rangle .
\end{align*}
Therefore we need the environment to be in a stationary state such
that $\mathrm{e}^{-i\Delta tH_{E}}\rho_{E}\mathrm{e}^{i\Delta tH_{E}}=\rho_{E}$
for all translation times $\Delta t$. Then

\begin{align*}
G_{n}(t_{1}+\Delta t,\dots,t_{n}+\Delta t)= & \frac{1}{2^{n-1}}\left\langle \left\{ B(t_{1}),\left\{ B(t_{2}),\left\{ \dots,B(t_{n})\right\} \dots\right\} \right\} \mathrm{e}^{-i\Delta tH_{E}}\rho_{E}\mathrm{e}^{i\Delta tH_{E}}\right\rangle \\
= & \frac{1}{2^{n-1}}\left\langle \left\{ B(t_{1}),\left\{ B(t_{2}),\left\{ \dots,B(t_{n})\right\} \dots\right\} \right\} \rho_{E}\right\rangle \\
= & G_{n}(t_{1},\dots,t_{n})\,.
\end{align*}
The cumulants thus have time translation symmetry if and only if the
environment is in a stationary state.

\subsection{Time reversal symmetry of the correlation functions and cumulants
of the environment \label{subapp:Time-reversal-environment}}

To satisfy the second condition $W_{n}(t_{1},\dots,t_{2n})=\left(\pm1\right)^{n}W_{n}(-t_{1},\dots,-t_{2n})$,
we need the cumulants to have time-reversal symmetry. We thus require
our full system with time-reversal symmetry. Time reversal symmetry
in simple terms involves reversing the direction of time in the equations
that describe the evolution of a system. If these equations remain
unchanged when time is reversed, the system is said to exhibit time
reversal symmetry. Quantum mechanics is unitary, i.e. the evolution
operator $U(t)$ is unitary and thus invertible $U^{\dagger}(t)=U^{-1}(t)$.
When a quantum system has time-reversal symmetry, this can be expressed
with the time reversal operator $T$, an anti-unitary operator that
commutes with the system Hamiltonian $[T,H]=0$ and such that $T^{-1}U(t)T=U^{-1}(t)$
\citep{Domingos1979}.

In our particular case, to have time-reversal symmetry there exists
some anti-unitarian operator $T=T_{S}\otimes T_{E}$ such that the
full Hamiltonian $H$ is $T-$invariant, i.e. $T^{-1}HT=H$. This
implies in particular that $T_{S}^{-1}S_{z}T_{S}=\pm S_{z}$, and
$T_{E}^{-1}BT_{E}=\pm B$, as the qubit-environment interaction $\propto S_{z}B$
must be $T-$invariant. We call the cases when $B$ commutes with
the time reversal operator $[B,T_{\text{env}}]=0$ of the electric-type,
as it is the case when $B$ represents gate charges, voltages, electric
fields, polarizations, etc. Analogously, we call the case when $B$
anticommutes with the time reversal operator $\{B,T_{\text{env}}\}=0$
of the magnetic-type, as it is the case when $B$ represents currents,
magnetic fields, magnetizations, etc. In our experimental realization
the noise operator is a magnetic field, therefore we performed experiments
in a magnetic-type case.

The time reversal operator $T_{E}$ must satisfy $T_{E}^{2}=\pm1$
\citep{Domingos1979}, therefore $T_{E}^{-1}=\pm T_{E}$. This in
turn implies
\begin{align*}
T_{E}^{-1}U_{E}(t)T_{E} & =U_{E}^{\dagger}(t)\,,\\
T_{E}^{-1}U_{E}^{\dagger}(t)T_{E} & =U_{E}(t)\,,\\
T_{E}U_{E}(t)T_{E}^{-1} & =U_{E}^{\dagger}(t)\,,\\
T_{E}U_{E}^{\dagger}(t)T_{E}^{-1} & =U_{E}(t)\,,
\end{align*}
where $U_{E}(t)=\mathrm{e}^{-iH_{E}t}$ is the environmental evolution
operator.  This thus implies that $T_{E}^{-1}B(t)T_{E}=T_{E}^{-1}U_{E}^{\dagger}(t)BU_{E}(t)T_{E}=\pm B(-t)$.

Based on this properties of the time-reversal operator, we analyze
the behavior of the cumulants under the time-reversal. As the cumulants
are written in terms of the environment correlation functions $G_{n}$
of Eq. (\ref{eq:def-G}), for the cumulants to have time-reversal
symmetry it is necessary and sufficient for the correlation functions
to have time reversal symmetry $G_{n}(t_{1},\dots,t_{n})=(\pm1)^{n}G_{n}(-t_{1},\dots,-t_{n})$.
Without loss of generality we consider $t_{1}\leq t_{2}\leq\dots\leq t_{n}$,
and thus $-t_{n}\leq-t_{n-1}\leq\dots\leq-t_{1}$. Therefore 
\[
G_{n}(-t_{1},\dots,-t_{n})=\frac{(\pm1)^{n}}{2^{n-1}}\left\langle \left\{ B(-t_{n}),\left\{ B(-t_{n-1}),\left\{ \dots,B(-t_{1})\right\} \dots\right\} \right\} \rho_{E}\right\rangle \,.
\]
We can now achieve to this expression using $T_{E}$ as 
\[
G_{n}(-t_{1},\dots,-t_{n})=\frac{(\pm1)^{n}}{2^{n-1}}\left\langle \left\{ T_{E}^{-1}B(t_{n})T_{E},\left\{ T_{E}^{-1}B(t_{n-1})T_{E},\left\{ \dots,T_{E}^{-1}B(t_{1})T_{E}\right\} \dots\right\} \right\} \rho_{E}\right\rangle \,.
\]
The product of the operators $T_{E}$ and $T_{E}^{-1}$ is the identity
and we get 
\[
G_{n}(-t_{1},\dots,-t_{n})=\frac{(\pm1)^{n}}{2^{n-1}}\left\langle \left\{ B(t_{n}),\left\{ B(t_{n-1}),\left\{ \dots,B(t_{1})\right\} \dots\right\} \right\} T_{E}\rho_{E}T_{E}^{-1}\right\rangle \,.
\]
We thus obtain that for the cumulants to have time-reversal symmetry,
we need the environmental state $\rho_{E}$ to be $T_{E}-$invariant,
i.e. $T_{E}\rho_{E}T_{E}^{-1}=\rho_{E}$. If we assume this, we obtain
\begin{equation}
G_{n}(-t_{1},\dots,-t_{n})=\frac{(\pm1)^{n}}{2^{n-1}}\left\langle \left\{ B(t_{n}),\left\{ B(t_{n-1}),\left\{ \dots,B(t_{1})\right\} \dots\right\} \right\} \rho_{E}\right\rangle .\label{eq:T-G}
\end{equation}
This differs from $(\pm1)^{n}G_{n}(t_{1},\dots,t_{n})$ in the ordering
of the noise operators in the anticommutators. For $G_{1}$ and $G_{2}$,
Eq. (\ref{eq:T-G}) becomes $G_{1}(t_{1})=\pm G_{1}(-t_{1})$ and
$G_{2}(t_{1},t_{2})=G_{2}(-t_{1},-t_{2})$. However from $n=3$ and
forth, $G_{n}(t_{1},\dots,t_{n})$ differs from $(\pm1)^{n}G_{n}(-t_{1},\dots,-t_{n})$.

For example for $t_{1}\leq t_{2}\leq t_{3}$, 
\begin{multline*}
G_{3}(t_{1},t_{2},t_{3})=\frac{1}{4}\left\langle \left\{ B(t_{1}),\left\{ B(t_{2}),B(t_{3})\right\} \right\} \rho_{E}\right\rangle \\
=\frac{1}{4}\left\langle \left[B(t_{1})B(t_{2})B(t_{3})+B(t_{1})B(t_{3})B(t_{2})+B(t_{2})B(t_{3})B(t_{1})+B(t_{3})B(t_{2})B(t_{1})\right]\rho_{E}\right\rangle \,,
\end{multline*}
 and 
\begin{multline*}
\pm G_{n}(-t_{1},-t_{2},-t_{3})=\frac{1}{4}\left\langle \left\{ B(t_{3}),\left\{ B(t_{2}),B(t_{1})\right\} \right\} \rho_{E}\right\rangle \\
=\frac{1}{4}\left\langle \left[B(t_{3})B(t_{2})B(t_{1})+B(t_{3})B(t_{1})B(t_{2})+B(t_{2})B(t_{1})B(t_{3})+B(t_{1})B(t_{2})B(t_{3})\right]\rho_{E}\right\rangle \,.
\end{multline*}
Therefore
\begin{align*}
G_{3}(t_{1},t_{2},t_{3}) & \mp G_{n}(-t_{1},-t_{2},-t_{3})=\\
= & \frac{1}{4}\left\langle \left[B(t_{1})B(t_{3})B(t_{2})+B(t_{2})B(t_{3})B(t_{1})-B(t_{3})B(t_{1})B(t_{2})-B(t_{2})B(t_{1})B(t_{3})\right]\rho_{E}\right\rangle \\
= & \frac{1}{4}\left\langle \left\{ \left[B(t_{1})B(t_{3})-B(t_{3})B(t_{1})\right]B(t_{2})+B(t_{2})\left[B(t_{3})B(t_{1})-B(t_{1})B(t_{3})\right]\right\} \rho_{E}\right\rangle \\
= & \frac{1}{4}\left\langle \left\{ \left[B(t_{1}),B(t_{3})\right]B(t_{2})+B(t_{2})\left[B(t_{3}),B(t_{1})\right]\right\} \rho_{E}\right\rangle \\
= & \frac{1}{4}\left\langle \left[\left[B(t_{1}),B(t_{3})\right],B(t_{2})\right]\rho_{E}\right\rangle \,.
\end{align*}
This thus shows an example on how the time reversal symmetry of higher
order correlation functions is broken and the breaking is proportional
to commutators of the noise operator at different times.

Nevertheless, if the environment is classical or it is at the high
temperature limit, one can make a semiclassical approximation of the
noise operators and replace 
\[
\left\{ B(t_{1}),B(t_{2})\right\} \sim2B(t_{1})B(t_{2})\sim2B(t_{2})B(t_{1})\,.
\]
In these cases, the correlation functions become 
\[
G_{n}(t_{1},\dots,t_{n})=\left\langle B(t_{1})B(t_{2})\dots B(t_{n})\rho_{E}\right\rangle =(\pm1)^{n}G_{n}(-t_{1},\dots,-t_{n}),
\]
and therefore also the cumulants satisfy 
\[
W_{n}(t_{1},\dots,t_{n})=(\pm1)^{n}W_{n}(-t_{1},\dots,-t_{n})\,,
\]
thus satisfying the time-reversal symmetry in the cumulants.

When the environment is Gaussian, either quantum or classical, the
only non-zero cumulants are $W_{1}$ and $W_{2}$. Since $G_{1}(t_{1})=\pm G_{1}(-t_{1})$
and $G_{2}(t_{1},t_{2})=G_{2}(-t_{1},-t_{2})$, then $W_{1}(t_{1})=\pm W_{1}(-t_{1})$
and $W_{2}(t_{1},t_{2})=W_{2}(-t_{1},-t_{2})$. Analogously, when
the qubit is weakly coupled to its environment, then the phase is
dominated by $W_{1}$ and the decay by $W_{2}$, and the environment
can be well approximated as a Gaussian environment even if it is a
quantum environment.

The cumulants thus only have time-reversal symmetry when the contributions
of noise operators commute at different times, if the environment
is quantum and non-Gaussian in the sense that is described with cumulants
of order higher than 2.

Summarizing, the results presented here mean that the time-reversal
symmetry is only attained when all the relevant cumulants to the qubit-probe
dephasing satisfy 
\begin{equation}
W_{n}(t_{1},\dots,t_{n})=(\pm1)^{n}W_{n}(-t_{1},\dots,-t_{n})\,.\label{eq:TW-de-W}
\end{equation}
Instead, the time-reversal symmetry is broken when the environment
is quantum, at low temperature, non-Gaussian, and strongly coupled
to the qubit.

\subsection{Time reversal symmetry of the decoherence factor \label{subapp:Time-reversal-qubit}}

When the environment is in a stationary state, and satisfies Eq.
(\ref{eq:TW-de-W}), this means that
\[
\mathcal{J}_{f_{T}}=-\sum_{n}(\pm1)^{n}\frac{i^{n}}{n!}\int_{0}^{T_{s}}\mathrm{d}t_{1}\dots\int_{0}^{T_{s}}\mathrm{d}t_{n}\,f(t_{1})\dots f(t_{n})W_{n}(t_{1},\dots,t_{n})\,,
\]
or, equivalently, 
\begin{align*}
\text{Re}\left(\mathcal{J}_{f_{T}}\right)= & \text{Re}\left(\mathcal{J}_{f_{T}}\right)\,,\\
\text{Im}\left(\mathcal{J}_{f_{T}}\right)= & \pm\text{Im}\left(\mathcal{J}_{f_{T}}\right)\,,
\end{align*}
where $\pm=+$ for electric-type noise and $\pm=-$ for magnetic-type
noise. This means that for electric-type noise $M_{f_{T}}=M_{f}$,
and for magnetic noise $M_{f_{T}}=M_{f}^{*}$.

\section{Filter out of stationary environmental noise sources by the SENSIT
contrast\label{sec:Filter-out_stationary}}

We here demonstrate the selective property of the SENSIT contrast
to filter in environmental sources that leads to the breaking of the
time reversed symmetry of the qubit-control, i.e. nonstationary or
quantum non-Gaussian noise sources. The SENSIT control filters out
the noise sources that induce decoherence on the qubit-probe, but
do not induce a breaking in the control time-reversal symmetry. Formally,
we consider an environment containing two separated subsystems $a$
and $b$, such that $a$ is either out of equilibrium inducing non-stationary
noise sources or it is quantum non-Gaussian and $b$ is an environment
that is stationary and effectively Gaussian or classical. This means
both $a$ and $b$ induce decoherence on the qubit-probe, but only
$a$ induces time-reversal symmetry breaking on the control. We write
the environment Hilbert space as $\mathcal{H}_{E}=\mathcal{H}_{a}\otimes\mathcal{H}_{b}$,
the noise operator $B=B_{a}\otimes\mathbb{I}_{b}+\mathbb{I}_{a}\otimes B_{b}$,
the environmental Hamiltonian $H_{E}=H_{a}\otimes\mathbb{I}_{b}+\mathbb{I}_{a}\otimes H_{b}$
and the initial environmental state as $\rho_{E}=\rho_{a}\otimes\rho_{b}$.
The environments have thus an independent initial state, evolve independently
and each of them adds a separated noise contribution to the qubit-probe.
For example, in the experiments we carried out on this work, the environment
$a$ would be the spin network of $^{1}H$ nuclei, while $b$ would
be any other sources of decoherence, e.g. the electrons and phonons
of the sample, cosmic rays and radiation going trough it, the current
fluctuations in the magnet, the noise induced by electrical installations,
etc. Since the environmental subsystems do not interact with each
other, the evolution of the noise operator with respect to the environmental
Hamiltonian is $B(t)=B_{a}(t)\otimes\mathbb{I}_{b}+\mathbb{I}_{a}\otimes B_{b}(t)$,
where $B_{i}(t)=\mathrm{e}^{iH_{i}t}B_{i}\mathrm{e}^{-iH_{i}t}$ for
$i=a,b$. The evolution operator $\mathcal{T}\mathrm{e}^{\frac{i}{2}\int_{0}^{T_{s}}\mathrm{d}t\,f(t)B(t)}$
in Eq. (\ref{eq:Generador_Gs}) is also separable
\[
\mathcal{T}\mathrm{e}^{\frac{i}{2}\int_{0}^{T_{s}}\mathrm{d}t\,f(t)B(t)}=\mathcal{T}\mathrm{e}^{\frac{i}{2}\int_{0}^{T_{s}}\mathrm{d}t\,f(t)B_{a}(t)}\otimes\mathcal{T}\mathrm{e}^{\frac{i}{2}\int_{0}^{T_{s}}\mathrm{d}t\,f(t)B_{b}(t)}.
\]
The decay of the qubit magnetization thus becomes 
\[
M=\left\langle \left(\mathcal{T}\mathrm{e}^{\frac{i}{2}\int_{0}^{T_{s}}\mathrm{d}t\,f(t)B_{a}(t)}\right)\rho_{a}\left(\mathcal{T}\mathrm{e}^{-\frac{i}{2}\int_{0}^{T_{s}}\mathrm{d}t\,f(t)B_{a}(t)}\right)^{\dagger}\right\rangle \times\left\langle \left(\mathcal{T}\mathrm{e}^{\frac{i}{2}\int_{0}^{T_{s}}\mathrm{d}t\,f(t)B_{b}(t)}\right)\rho_{b}\left(\mathcal{T}\mathrm{e}^{-\frac{i}{2}\int_{0}^{T_{s}}\mathrm{d}t\,f(t)B_{b}(t)}\right)^{\dagger}\right\rangle \,.
\]
We now define the cumulants $W_{n}^{a/b}$ of the subsystems as the
cumulants of the respective noise operators $B^{a/b}(t)$. We can
thus write the contributions to the decay in terms of 
\[
\left\langle \left(\mathcal{T}\mathrm{e}^{\frac{i}{2}\int_{0}^{T_{s}}\mathrm{d}t\,f(t)B_{i}(t)}\right)\rho_{i}\left(\mathcal{T}\mathrm{e}^{-\frac{i}{2}\int_{0}^{T_{s}}\mathrm{d}t\,f(t)B_{i}(t)}\right)^{\dagger}\right\rangle =\exp\left[\sum_{n}\frac{\left(i\right)^{n}}{n!}\int_{0}^{T_{s}}\mathrm{d}t_{1}\dots\int_{0}^{T_{s}}\mathrm{d}t_{n}\,f(t_{1})\dots f(t_{n})W_{n}^{i}(t_{1},\dots,t_{n})\,,\right]
\]
and we obtain the decoherence factor as 
\[
\mathcal{J}=-\sum_{n}\frac{\left(i\right)^{n}}{n!}\int_{0}^{T_{s}}\mathrm{d}t_{1}\dots\int_{0}^{T_{s}}\mathrm{d}t_{n}\,f(t_{1})\dots f(t_{n})\left[W_{n}^{a}(t_{1},\dots,t_{n})+W_{n}^{b}(t_{1},\dots,t_{n})\right]\,,
\]
and compare with Eq. (\ref{eq:J-de-W}), we obtain that the cumulants
of the environment $W_{n}=W_{n}^{a}+W_{n}^{b}$ are the sum of the
cumulants of each subsystem. Since the environment $b$ is neither
non-stationary nor quantum non-Gaussian, the cumulants $W_{n}^{b}$
are symmetric under time reversed control. Therefore, the SENSIT contrast
is 
\[
\Delta\mathcal{J}=-\sum_{n}\frac{i^{n}}{\left(n\right)!}\int_{0}^{T_{s}}\mathrm{d}t_{1}\dots\int_{0}^{T_{s}}\mathrm{d}t_{n}\,f(t_{1})\dots f(t_{n})\,\Delta W_{n}^{a}(t_{1},\dots,t_{n})\,,
\]
where $\Delta W_{n}^{a}(t_{1},\dots,t_{2n})=W_{n}^{a}(t_{1},\dots,t_{2n})-W_{n}^{a}(T_{s}-t_{1},\dots,T_{s}-t_{2n})$.
This thus shows how the SENSIT contrast is independent of all properties
of $b$, and is in fact the same as what would be obtained if $a$
were the entire environment of the qubit-probe. This shows that SENSIT
is filtering out all noise sources leading to time-reversal symmetry
on the qubit control, while maintaining sensitivity to those out of
equilibrium or quantum non-Gaussian.

\section{SENSIT contrast with an environment near to a stationary state\label{sec:SENSIT-contrast-nearEquilibrium}}

To give an example on how the distance to equilibrium is encoded in
the SENSIT contrast, we consider a simple but general case where the
environment is near to a stationary state $\rho_{E}^{(0)}$. We consider
that quantum non-Gaussian contributions to the cumulants are negligible.
That is, $\rho_{E}^{(0)}$ is a state where the cumulants have time
reversal symmetry in the qubit-probe control $\Delta W_{n}=0$. We
consider the environmental sate to be $\rho_{E}=\rho_{E}^{(0)}+\epsilon\rho_{E}^{(1)}$,
where $\rho_{E}^{(1)}$ is a constant perturbation to the environmental
state with $\epsilon$ quantifying the distance to equilibrium. Under
these conditions, the cumulants can be expanded in the distance $\epsilon$
as $W_{n}=W_{n}^{(0)}+\epsilon W_{n}^{(1)}+\mathcal{O}\left(\epsilon^{2}\right)$,
where $W_{n}^{(0)}$ is the $n$-th cumulant for the environmental
state $\rho_{E}^{(0)}$, and $W_{n}^{(1)}$ depends on both environmental
contributions $\rho_{E}^{(0)}$ and $\rho_{E}^{(1)}$. For example,
for $n=2$ 
\[
W_{2}^{(1)}(t_{1},t_{2})=\left\langle \left\{ B(t_{1}),B(t_{2})\right\} \rho_{E}^{(1)}\right\rangle -\left\langle B(t_{1})\rho_{E}^{(1)}\right\rangle \left\langle B(t_{2})\rho_{E}^{(0)}\right\rangle -\left\langle B(t_{1})\rho_{E}^{(0)}\right\rangle \left\langle B(t_{2})\rho_{E}^{(1)}\right\rangle \,.
\]
Since we consider that $\rho_{E}^{(0)}$ has time reversal symmetry
in the qubit-probe control, the main contributions to $\Delta W_{n}$
will be linear in $\epsilon$ and due to the term 
\[
\Delta W_{n}^{(1)}(t_{1},\dots,t_{n})=W_{n}^{(1)}(t_{1},\dots,t_{n})-\left(\pm1\right)^{n}W_{n}^{(1)}(T_{s}-t_{1},\dots,T_{s}-t_{n})\,.
\]
The measured SENSIT contrast is thus
\[
\Delta\mathcal{J}=-\epsilon\sum_{n}\frac{i^{n}}{\left(n\right)!}\int_{0}^{T_{s}}\mathrm{d}t_{1}\dots\int_{0}^{T_{s}}\mathrm{d}t_{n}\,f(t_{1})\dots f(t_{n})\,\Delta W_{n}^{(1)}(t_{1},\dots,t_{n})\,+\mathcal{O}\left(\epsilon^{2}\right),
\]
manifesting that when the environment is near to a state that satisfies
the conditions for time-reversal symmetry in the qubit-probe control,
the SENSIT contrast is proportional to the parameter $\epsilon$ that
quantifies the distance to the stationary environmental state.

\section{SENSIT Contrast of a Quenched Ornstein-Uhlenbeck Process\label{sec:SENSIT-Contrast-quenchedOU}}

We consider the paradigmatic example of a quenched OU process for
the noise operator fluctuations, with a self-correlation time $\tau$
and the standard deviation $\sigma_{0}$ at equilibrium, i.e. at long
times. This stochastic process is Gaussian with zero mean, thus all
cumulants with $n\neq2$ are null $W_{n}=0$. The quench is generated
by an out-of-equilibrium initial condition, corresponding to a Gaussian
distribution with a standard deviation different from the one at equilibrium
$\sigma\neq\sigma_{0}$. The self-correlation function of the quenched
process is
\begin{eqnarray*}
W_{2}\left(t_{1},t_{2}\right) & = & W_{2}^{\text{Eq}}\left(t_{1},t_{2}\right)+W_{2}^{\text{Q}}\left(t_{1},t_{2}\right)\,,
\end{eqnarray*}
where 
\begin{eqnarray*}
W_{2}^{\text{Eq}}\left(t_{1},t_{2}\right) & = & \sigma_{0}\mathrm{e}^{-\frac{\left|t_{1}-t_{2}\right|}{\tau}\,}
\end{eqnarray*}
and 
\begin{eqnarray*}
W_{2}^{\text{Q}}\left(t_{1},t_{2}\right) & = & \left(\sigma-\sigma_{0}\right)\mathrm{e}^{-\frac{t_{1}+t_{2}}{\tau}}\,
\end{eqnarray*}
are the equilibrium and quench contributions to the correlation function
\citep{Kuffer2022}. The SENSIT contrast measured from a qubit-probe
coupled to this noise process is $\text{Re}\Delta\mathcal{J}=-\int_{0}^{T_{s}}\mathrm{d}t_{1}f(t_{1})\int_{0}^{T_{s}}\mathrm{d}t_{2}f(t_{2})\left[W_{2}^{\text{Q}}\left(t_{1},t_{2}\right)-W_{2}^{\text{Q}}\left(T_{s}-t_{1},T_{s}-t_{2}\right)\right]$.
Since $W_{2}^{\text{Eq}}\left(t_{1},t_{2}\right)=W_{2}^{\text{Eq}}\left(T_{s}-t_{1},T_{s}-t_{2}\right)$,
the equilibrium correlation function does not contribute to the SENSIT
contrast. After replacing the cumulants, the SENSIT contrast can be
written as $\text{Re}\Delta\mathcal{J}=\left(\sigma-\sigma_{0}\right)\Sigma\left[f\right]$,
with 
\[
\Sigma\left[f\right]=-\frac{1}{2}\left[\left(\int_{0}^{T_{s}}\mathrm{d}tf(t)\,\mathrm{e}^{-\frac{t}{\tau}}\right)^{2}-\left(\int_{0}^{T_{s}}\mathrm{d}tf(t)\mathrm{e}^{-\frac{T_{s}-t}{\tau}}\right)^{2}\right]\,.
\]
Note that the SENSIT contrast is proportional to $\left(\sigma-\sigma_{0}\right)$,
i.e. a variance distance defining how far from equilibrium the initial
state of the environment is, weighted by the term $\Sigma\left[f\right]$
that measures the ability of the chosen control modulation function
$f$ to detect the time-reversal symmetry breaking. This thus sets
a paradigmatic example about how the SENSIT contrast is proportional
to $\left(\sigma-\sigma_{0}\right)$ quantifying the distance from
the initial state of the environment to its stationary state at equilibrium.

\section{Solid-state NMR Quantum Simulations \label{met:Quantum-Simulations}}

\subsection{Experimental setup and system\label{submet:setup_experimental}}

The quantum simulations were performed with solid-state NMR experiments
on a Bruker Avance III HD 9.4T WB spectrometer with a $^{1}\text{H}$
resonance frequency of $400.15$ MHz and a $^{13}\text{C}$ resonance
frequency of $100.61$ MHz. We used the nuclear spins of a polycrystalline
adamantane $C_{10}H_{16}$ sample to set up the qubit-probe and its
environment. Most hydrogen nuclei (98.93\%) in the sample are protons,
with a spin $\nicefrac{1}{2}$, while only approximately $1.07\%$
of the carbons are $^{13}C$, with spin $1/2$, the remaining carbons
has no magnetic moment. The low concentration of $^{13}$C allows
to neglect the interaction between them as the interaction with the
hydrogens is dominant. We therefore consider the sample as an ensemble
of $^{13}C$ qubit-probes that interact with the protons near them,
considered as the environment.

The experiments are carried out at a high magnetic field, as the Zeeman
energy is $\gtrsim10^{5}$ times stronger that the spin-spin interactions.
Thus only the secular part of the internuclear dipolar Hamiltonian
contributes to the dynamics \citep{slichter_principles_1996}. Therefore
the full Hamiltonian of the system is
\begin{equation}
H=H_{S}+H_{E}+H_{SE}\,,\label{eq:H-total-Schrodinger}
\end{equation}
where the qubit-probe (carbon) considered as our system $S$, the
environment (protons) and the system-environment interaction Hamiltonians
are

\begin{subequations}
\begin{align*}
H_{S} & =\gamma_{C}B_{0}S_{z}\otimes\mathbb{I}_{E}+\text{control}\,,\\
H_{E} & =\gamma_{H}B_{0}\mathbb{I}_{S}\otimes I_{z}+\sum_{i\neq j}d_{ij}^{H}\left(2I_{z}^{i}I_{z}^{j}-I_{x}^{i}I_{x}^{j}-I_{y}^{i}I_{y}^{j}\right)+\text{control}\,,\\
H_{SE} & =S_{z}\sum_{i}d_{i}I_{z}^{i}=S_{z}B\,,
\end{align*}
respectively.

\end{subequations}Here, the sums run over the environmental spins
(the protons), $\gamma_{C}$ and $\gamma_{H}$ are the gyromagnetic
ratios of the $^{13}C$ and proton, respectively, $B_{0}$ is the
static field applied in the $z$-direction, the spin operators $\mathbf{S}$
is the qubit-probe ($^{13}C$ spin) and $\mathbf{I}^{i}$ are the
environmental spins with $\mathbf{I}=\sum_{i}\mathbf{I}^{i}$ the
total spin moment of the environment, $d_{ij}^{H}$ are the dipolar
couplings between the environmental spins $i$ and $j$, $d_{i}$
is the dipolar coupling between the qubit-probe and the $i-$th spin
of the environment. The noise operator is $B=\sum_{i}d_{i}I_{z}^{i}$
that represents the field that the qubit-probe experiences due to
the environment.

In the interaction picture with respect to the Zeeman interactions,
the Hamiltonians become \citep{slichter_principles_1996}
\begin{align*}
H_{S} & =\text{control}\,,\\
H_{E} & =\sum_{i\neq j}d_{ij}^{H}\left(2I_{z}^{i}I_{z}^{j}-I_{x}^{i}I_{x}^{j}-I_{y}^{i}I_{y}^{j}\right)+\text{control}\,.\\
H_{SE} & =S_{z}\sum_{i}d_{i}I_{z}^{i}=S_{z}B\,.
\end{align*}
We utilize control over the environment solely to create the nonstationary
initial state. Subsequently, during the quantum simulations of the
SENSIT protocol, only the qubit-probe is controlled. For our experiments,
we applied on-resonance $\pi$-pulses with the qubit-probe. In the
interaction picture with respect to this control, i.e. the toggling
frame \citep{abragam_principles_1983}, the Hamiltonian is
\begin{align*}
H_{S} & =0\,,\\
H_{E} & =\sum_{i\neq j}d_{ij}^{H}\left(2I_{z}^{i}I_{z}^{j}-I_{x}^{i}I_{x}^{j}-I_{y}^{i}I_{y}^{j}\right),\\
H_{SE} & =f(t)S_{z}\sum_{i}d_{i}I_{z}^{i}=f(t)S_{z}B\,,
\end{align*}
where the function $f(t)$, switches its signs whenever a $\pi$-pulse
is applied \citep{Suter2016}.

These Hamiltonians have time-reversal symmetry. To write this explicitly,
we use the representation $I_{\alpha}^{i}=\frac{1}{2}\sigma_{\alpha}^{i}$,
$S_{z}=\frac{1}{2}\sigma_{\alpha}^{S}$, where $\sigma_{\alpha}^{i}$
are Pauli matrices that act on the space of the $i$-th environment
spin and $\sigma_{\alpha}^{S}$ are Pauli matrices that act on the
qubit space. The time reversal operator is 
\begin{equation}
T=\sigma_{y}^{S}\prod_{i}\sigma_{y}^{\text{i}}K\,,\label{eq:def_time_reversal}
\end{equation}
where $K$ is the complex conjugation operator \citep{Domingos1979}.
Since the system is at high temperature, the contribution of noncommutation
terms to the SENSIT contrast will be negligible, thus any measured
SENSIT contrast is due to asymmetry generated by the environment initial
state. In particular, if the state commutes with the environmental
Hamiltonian and has time-reversal symmetry, then the system will have
time-reversal symmetry in the control functions.

To finally obtain a Hamiltonian with the form of Eq. (\ref{eq:H_I(t)})
in the main text, we switch to an interaction picture with respect
to the environmental Hamiltonian
\[
H(t)=H_{SE}(t)=f(t)S_{z}B(t),
\]
where $B(t)=\mathrm{e}^{iH_{E}t}B\mathrm{e}^{-iH_{E}t}$ is the time-dependent
noise operator. This is the experimental implementation of Eq. (\ref{eq:H_I(t)})
of the main text in our quantum simulations.

\subsection{Out-of-equilibrium environmental-state preparation\label{submet:state-preparation}\label{submet:SDR}}

\begin{figure}
\includegraphics[width=1\textwidth]{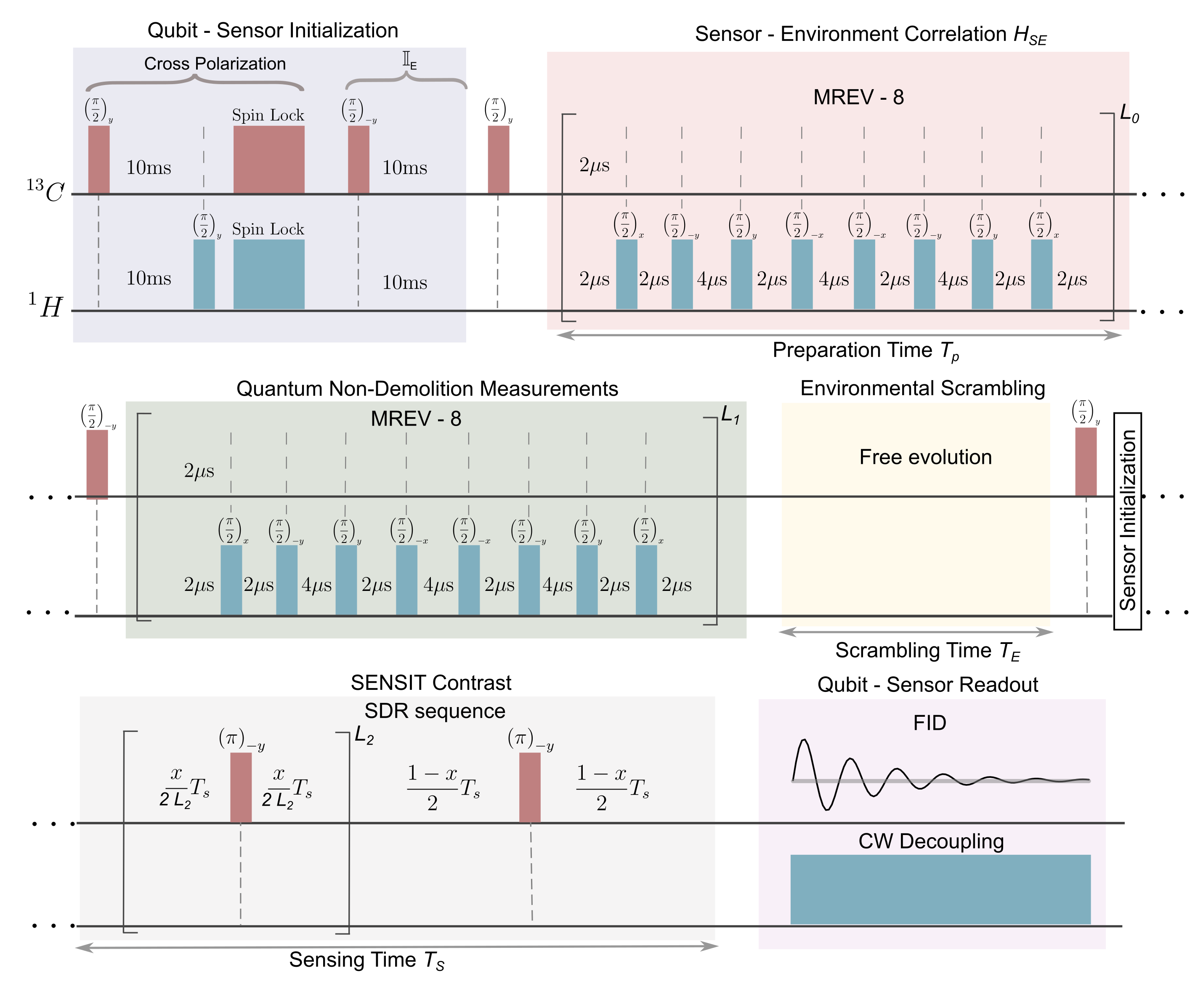}\caption{\label{fig:Secuencia-Pulsos-Completa}Pulse sequence used for the
SENSIT experiments. The different blocks represents different conceptual
steps. First, cross polarization is used to initialize the system
and increase the signal-to-noise ratio. Then, the $10$ms delay time
ensures that all qubit-environment correlations and environmental
initial state are erased due to $T_{2}$ relaxation before further
control. This is then followed by a $\left(\nicefrac{\pi}{2}\right)_{y}$
pulse, to put the qubit-probe state in-plane. Then $L_{0}$ cycles
of the MREV-8 sequence are implemented to correlate the qubit-probe
and the environment and a $\left(\nicefrac{\pi}{2}\right)_{-y}$ pulse
returns the qubit-probe state to the $z$ direction. The preparation
time $T_{p}$ is the total duration of the $L_{0}$ MREV-8 cycles.
After that, the MREV-8 sequence is used again to mimic a projective
measurement on the qubit-probe and turn the correlated state into
a state where the qubit and environment are uncorrelated, but with
the environment out of equilibrium. To ensure the completeness of
the projective measurement, we attain a stationary state by varying
$L_{1}$. Specifically, we set $L_{1}=42$ because for larger values,
the qubit-probe state remained unchanged despite alterations in $L_{1}$.
The environment then is scrambled by letting it evolve freely during
an environmental scrambling time $T_{E}$. The qubit-probe state is
then put in plane with a $\left(\nicefrac{\pi}{2}\right)_{y}$ pulse,
and the SDR sequence is applied during the sensing time $T_{s}$ to
sense the environment via the readout of the qubit-probe state. During
this readout, the environment is decoupled for improving the signal-to-noise
ratio using continuous wave decoupling. To obtain the SENSIT contrast
the experiment is repeated exchanging the SDR sequence by the TSDR
sequence.}
\end{figure}

In this section we describe the pulse sequence used in the experiments
described in Fig. \ref{fig:Secuencia-Pulsos-Completa}. For the quantum
simulations, we first need to prepare an out-of-equilibrium environment
state of the form 
\[
\rho_{0}=S_{x}\otimes\rho_{E}\,,
\]
with $\left[\rho_{E},H_{E}\right]\neq0$. Before this preparation,
the initial state of the full system is in a Boltzmann thermal equilibrium
state. Since the experiments were conducted at room temperature, and
the Zeeman energy is $\lesssim15000$ times lower than the thermal
energy, the state is described in the high temperature limit \citep{slichter_principles_1996}.
Therefore, the initial state is $\rho\simeq\frac{1}{\text{tr}[\mathbb{I}]}\left(\mathbb{I}-\beta H\right)$.
As the Zeeman coupling to the external magnetic field is dominant
over the dipolar coupling between nuclei, the state is further simplified
$\rho\simeq\frac{1}{\text{tr}[\mathbb{I}]}\left[\mathbb{I}-\beta\gamma_{C}B_{0}S_{z}\otimes\mathbb{I}_{E}-\beta\gamma_{H}B_{0}\mathbb{I}_{S}\otimes I_{z}\right]$.
The initial step in the experiments involves performing cross-polarization
between the environment (the protons) and the qubit (the carbons)
just to increase the nuclear polarization of the carbons, and thus
improve the signal-to-noise ratio of the qubit-probe signal by a factor
of $\nicefrac{\gamma_{H}}{\gamma_{C}}\simeq3.97$. Then to remove
any possible correlation between the carbons and protons, we store
the qubit (carbon) magnetization in the $z$ direction and wait $10$ms
(time longer than $T_{2}$ of both spin species) to scramble out any
proton signal and carbon-proton correlations. This process is shown
in the Qubit-Sensor Initialization block of Fig. \ref{fig:Secuencia-Pulsos-Completa}.
The $\frac{\pi}{2}$ pulse durations we used for the protons and carbons
were $3.4\mu$s and $5.5\mu$s, respectively. The duration of the
carbon $\pi$ pulses was $11\mu$s. After this we apply a $\left(\nicefrac{\pi}{2}\right)_{y}$-pulse
on the qubit-probe around the y-axis, to obtain the state of the qubit-environment
system$\rho\simeq\frac{1}{\text{tr}[\mathbb{I}]}\left[\mathbb{I}-\beta\gamma_{H}B_{0}S_{x}\otimes\mathbb{I}_{E}\right]$.We
can thus write the state of our system just before the preparation
time period as $\rho=\left(\frac{1}{2}+\tilde{p}S_{x}\right)\otimes\frac{\mathbb{I}_{E}}{\text{tr}\left[\mathbb{I}_{E}\right]}$,
a separable state where the environment is at infinite temperature
and the qubit-probe is polarized in the $x$ direction. Here the polarization
of the qubit is $\tilde{p}=\beta\gamma_{H}B_{0}$. Hereafter, we drop
the term $\frac{1}{2}\otimes\frac{\mathbb{I}_{E}}{\text{tr}\left[\mathbb{I}_{E}\right]}$,
since it does not contribute to the qubit-probe signal, and we get
\begin{equation}
\rho=\tilde{p}S_{x}\otimes\frac{\mathbb{I}_{E}}{\text{tr}\left[\mathbb{I}_{E}\right]}\,,\label{eq:rho_termica_relevante}
\end{equation}
The environmental state $\frac{\mathbb{I}_{E}}{\text{tr}\left[\mathbb{I}_{E}\right]}$
is proportional to the identity and thus it is invariant under the
time reversal operation of Eq. (\ref{eq:def_time_reversal}), and
it also commutes with the environmental Hamiltonian. Since the experiments
are carried out at high temperature (See SI \ref{submet:setup_experimental}),
if the environmental state $\rho_{E}$ is stationary and invariant
under time reversal, then the conditions outlined in SI \ref{app:T-symmetry}
are satisfied, indicating that the system possesses time-reversal
symmetry in the qubit-probe control functions. We therefore predict
to observe a null SENSIT contrast when the environment is in this
state.

Then to turn this state into an out-of-equilibrium state, we perform
the following preparation step. We apply the MREV-8 pulse sequence
\citep{Rhim1973} during the preparation time $T_{p}$ on the environmental
spins to decouple them, i.e. cancel out the dipole-dipole interaction
between them thus making null the environmental Hamiltonian. We used
a MREV cycle time of $54,4\mu$s. This is shown in Fig. \ref{fig:Secuencia-Pulsos-Completa},
in the block labeled Sensor-Environment Correlation $H_{SE}$. The
effective full system Hamiltonian $H$ in the rotating frame of both
the qubit-probe and the environmental spins (i.e., in the interaction
picture with respect to the Zeeman Hamiltonians) is solely the interaction
Hamiltonian $H_{SE}$, with the field $B$ fixed in time as the evolution
stemming from the environmental Hamiltonian is refocused by the MREV
sequence
\[
H_{SE}=S_{z}\sum_{i}d_{i}I_{z}^{i}=S_{z}B\,.
\]
 During this step the qubit dephases due to interaction with the environment.
The full qubit-environment state becomes $\rho(T_{p})=\tilde{p}\mathrm{e}^{-iS_{z}BT_{p}}\left(\frac{1}{2}+\tilde{p}S_{x}\right)\otimes\frac{\mathbb{I}_{E}}{\text{tr}\left[\mathbb{I}_{E}\right]}\mathrm{e}^{iS_{z}BT_{p}}$,
or equivalently
\begin{equation}
\rho(T_{p})=\tilde{p}\left(S_{x}\otimes\frac{\cos\left(BT_{p}\right)}{\text{tr}\left[\mathbb{I}_{E}\right]}+S_{y}\otimes\frac{\sin\left(BT_{p}\right)}{\text{tr}\left[\mathbb{I}_{E}\right]}\right)\,.\label{eq:estado_correlacionado_inplane}
\end{equation}
As this state is not separable, we have created quantum correlations
between the environmental spins and the qubit-probe. This is simply
a free induction decay due to the constant but random field $B$ felt
by the qubit-probe.

We then proceeded to erase the qubit-environment correlations mimicking
a quantum non-demolition (QND) measurement on the qubit-probe, while
maintaining the non-equilibrium status of the environmental state
to attain a state like the one described in Eq. (\ref{eq:rho_inicial}).
To do this, we apply a $\left(\nicefrac{\pi}{2}\right)_{-y}$-pulse
around the $-y$ direction, and get the state
\[
\tilde{p}\left[S_{z}\otimes\frac{\cos\left(BT_{p}\right)}{\text{tr}\left[\mathbb{I}_{E}\right]}+S_{y}\otimes\frac{\sin\left(BT_{p}\right)}{\text{tr}\left[\mathbb{I}_{E}\right]}\right]\,.
\]
The first term satisfies $\left[S_{z}\otimes\frac{\cos\left(BT_{p}\right)}{\text{tr}\left[\mathbb{I}_{E}\right]},H_{SE}\right]=0$
and thus does not evolve, but the second one does evolve, and it will
be dephased --vanished-- due to the interaction with the environmental
field $B$. We thus apply the MREV-8 sequence for a sufficient duration
to dephase the second term, effectively simulating a quantum non-demolition
(QND) measurement on the qubit-probe state, where the state is projected
onto the subspace proportional to $S_{z}$. This is shown in the Quantum
Non-Demolition Measurements block in Fig. \ref{fig:Secuencia-Pulsos-Completa}.
As long as we do not refocus this dephasing with a time-reversion,
the qubit-probe signal we measure only comes from the state $\tilde{p}S_{z}\otimes\frac{\cos\left(BT_{p}\right)}{\text{tr}\left[\mathbb{I}_{E}\right]}$.

For the experiments described in Fig. \ref{fig:Scrambling_dependent}
of the main text, this is the step where environmental scrambling
is introduced. During the scrambling time $T_{E}$ a waiting period
is introduced without applying the MREV-8 sequence, as shown in Fig.
\ref{fig:Secuencia-Pulsos-Completa} within the Environmental Scrambling
block, so the environment Hamiltonian produces the information scrambling
on the environmental state. The created state is thus $\tilde{p}S_{z}\otimes\mathrm{e}^{-iH_{E}T_{E}}\frac{\cos\left(BT_{p}\right)}{\text{tr}\left[\mathbb{I}_{E}\right]}\mathrm{e}^{iH_{E}T_{E}}$.
This scrambling step is skipped for the experiments of Fig. \ref{fig:Preparation_dependent}.

Finally we apply a last $\left(\nicefrac{\pi}{2}\right)_{y}$-pulse
(Fig. \ref{fig:Secuencia-Pulsos-Completa}, Sensor Initialization),
and obtain when the scrambling time is included on the sequence
\[
\rho_{0}=\tilde{p}S_{x}\otimes\mathrm{e}^{-iH_{E}T_{E}}\frac{\cos\left(BT_{p}\right)}{\text{tr}\left[\mathbb{I}_{E}\right]}\mathrm{e}^{iH_{E}T_{E}}\,.
\]
This state is again separable, but it does not commute with the environment
Hamiltonian, as $\left[B,H_{E}\right]\neq0$, and thus it is now in
an out-of-equilibrium state that produces non-stationary noise fluctuations
on the qubit-probe. Thus, we rewrite this as
\[
\rho_{0}=pS_{x}\otimes\rho_{E}\,,
\]
where $p=\tilde{p}\frac{\text{tr}\left[\cos\left(BT_{p}\right)\right]}{\text{tr}\left[\mathbb{I}_{E}\right]}$
is the initial qubit-probe polarization for the sensing process, and
$\rho_{E}=\mathrm{e}^{-iH_{E}T_{E}}\frac{\cos\left(BT_{p}\right)}{\text{tr}\left[\cos\left(BT_{p}\right)\right]}\mathrm{e}^{iH_{E}T_{E}}$
is the out-of-equilibrium environmental density matrix. In experiments
of Fig. \ref{fig:Preparation_dependent}, without including the scrambling
time, the environmental state is $\rho_{E}=\frac{\cos\left(BT_{p}\right)}{\text{tr}\left[\cos\left(BT_{p}\right)\right]}$.
Again, since $\left[B,H_{E}\right]\neq0$ this state is nonstationary.
This shows how the initial out-of-equilibrium state is prepared in
our experiments.

The last step of the pulse sequence measures the SENSIT contrast.
In our experiments, we applied the selective dynamical recoupling
(SDR) sequence \citep{Smith2012,Alvarez2013a}, which is based on
a dynamical decoupling sequences that is asymmetric under time-reversal,
thus satisfying the requirements for obtaining nonzero SENSIT contrasts.
Figure \ref{fig:Secuencia-Pulsos-Completa} shows the SDR implementation
of the sequence in the SDR sequence block. We used $12$ pulses and
a total sensing time of $T_{s}=750\mu\text{s}$ for the experiments
of Fig. \ref{fig:Preparation_dependent} and \ref{fig:Scrambling_dependent}.
To measure the SENSIT contrast we also applied the the time reversed
version of SDR, the TSDR sequence, by reversing the order in which
the pulses and delays are applied. The SDR sequence interpolates continuously
a Hahn echo sequence of duration $T_{s}$ with a CPMG sequence consisting
of $N$ equidistant $\pi$ pulses between $t=0$ and $t=T_{s}$. It
consists of $N-1$ equidistant $\pi$ pulses between $t=0$ and $t=xT_{s}$
and a last $\pi$ pulse at $\frac{x+1}{2}T_{s}$. The dimensionless
parameter $x$ of the sequence interpolates the sequence between the
Hahn echo at $x=0$ and the CPMG sequence of $N$ pulses at $x=\frac{N-1}{N}$.
It can be interpreted as concatenating $N-1$ spin echoes of duration
$\frac{x}{N-1}T_{s}$, and a last single spin echo of duration $(1-x)T_{s}$.
For the time-reversed version TSDR , one first applies the single
spin echo of duration $(1-x)T_{s}$, and then concatenates the $N-1$
spin echoes of duration $\frac{x}{N-1}T_{s}$. The two different sequences
are shown side by side in Fig. \ref{fig:Preparation_dependent} of
the main text. Finally, the signal of the qubit-probe after the SDR
and TSDR sequences is measured under CW decoupling being applied to
the protons to increase the signal-to-noise ratio (Fig. \ref{fig:Secuencia-Pulsos-Completa},
Qubit-Sensor Readout).

The SENSIT contrast vanishes for $T_{p}=0$ as the system is in the
state described in Eq. (\ref{eq:rho_termica_relevante}). As the preparation
time $T_{p}$ increases, the state is driven further away from equilibrium,
leading to a growth in the nonstationary contributions to the cumulants.
Therefore, a growth of the SENSIT contrast is expected with increasing
$T_{p}$ , as shown in Fig. \ref{fig:Preparation_dependent}c of the
main text. If the scrambling time $T_{E}$ is included, we anticipate
a decrease in the SENSIT contrast with increasing $T_{E}$. This is
because the local information about the nonstationary state becomes
increasingly scrambled in nonlocal degrees of freedom that are not
accessible from the sensor qubit. This phenomenon is shown in Fig.
\ref{fig:Scrambling_dependent} of the main text.

\subsection{Quantifying the number $K$ of correlated spins in the Environment
due to initial state preparation\label{submet:K}}

\begin{figure}
\includegraphics[width=1\textwidth]{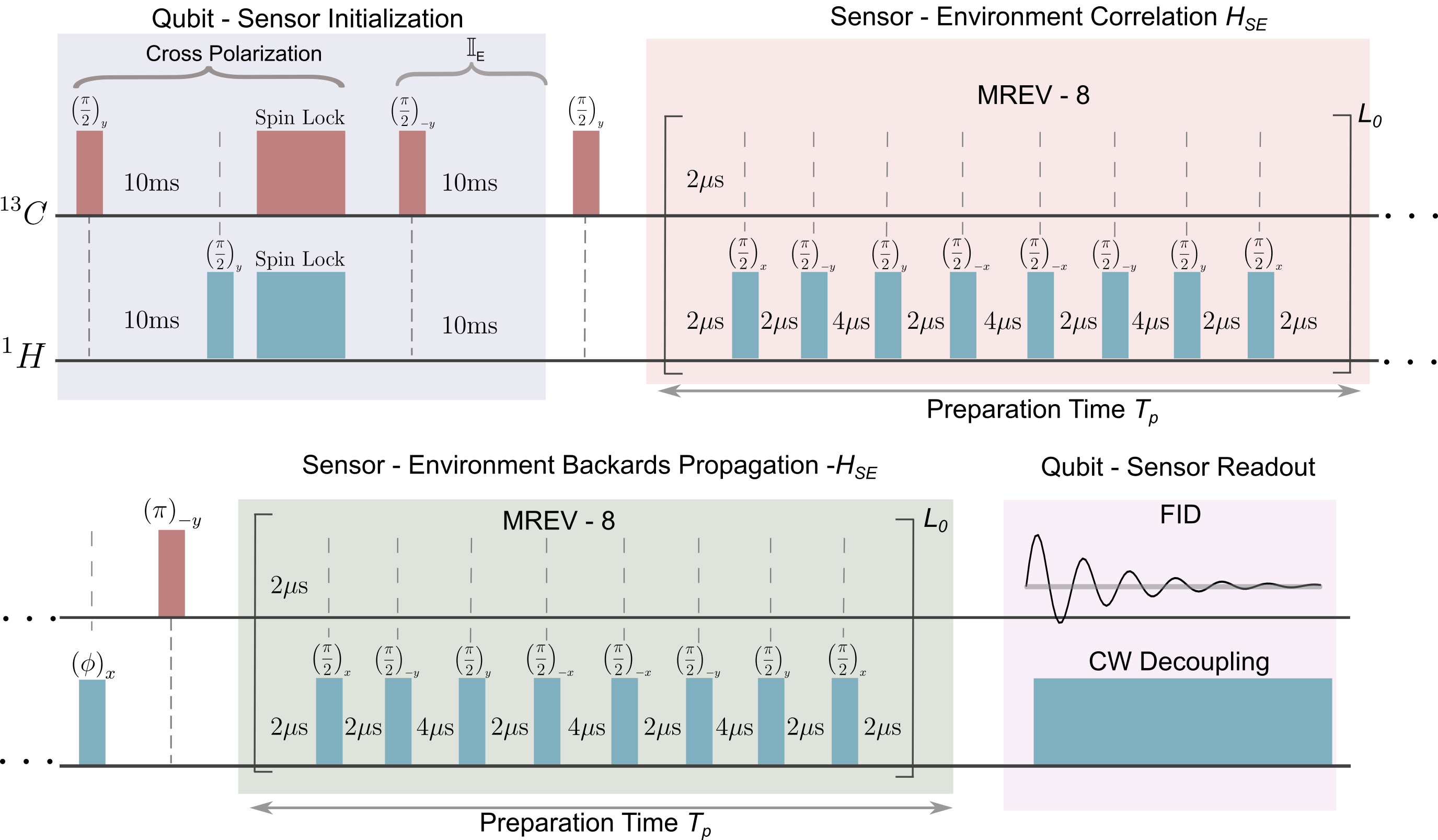}\caption{\label{fig:Secuencia-Pulsos-K}Pulse sequence used to measure the
number $K$ of environmental spins correlated to the qubit-sensor.
First, cross polarization is used to initialize the system and increase
the signal-to-noise ratio. Then, the $10$ms delay time ensures that
all qubit-environment correlations and environmental initial state
are erased due to $T_{2}$ relaxation before further control. This
is followed by a $\left(\nicefrac{\pi}{2}\right)_{y}$ pulse, putting
the qubit-probe state in plane. Then $L_{0}$ cycles of the MREV-8
sequence are implemented to correlate the qubit-probe and the environment,
as done when measuring the SENSIT contrast. After this, a $\left(\phi\right)_{x}$
pulse is applied on the environment to encode the formed correlations
\citep{Niknam2020}. Then an time-reversal echo is created by means
of a $\left(\pi\right)_{-y}$ pulse on the qubit-probe followed by
$L_{0}$ cycles of the MREV-8 sequence to produce an effective backwards
evolution. Finally the qubit-probe state is read out, while it is
decoupled from the environment with continuous wave irradiation for
improving the signal-to-noise ratio.}
\end{figure}

As we use the qubit-probe as a resource to create the out-of-equilibrium
state in the environment, by driven correlations between the qubit
and environmental spins, we measure the number of correlated environmental
spins as a measure of the out-of-equilibrium degree. We use the method
developed in \citep{Baum1985} to measure the multiple-quantum coherence
(MQC) spectrum of the density matrix state. Based on this approach
the second moment of the MQC spectrum gives an out-of-time-order commutator
$K$ that quantifies the number of correlated spins where the information
was scrambled from a local probe \citep{Alvarez2015,Dominguez2020,Dominguez2021,Niknam2020}
. Given that the nonstationary state post-preparation arises from
correlations established between the qubit-sensor and the environment,
the out-of-time-order commutator $K$ serves as an indicator of the
system's distance from equilibrium.

To measure the number of correlated spins $K$ we begin with the same
initialization as for the experiments to measure the SENSIT contrast:
A cross polarization followed by a dephasing of correlations (see
Fig. \ref{fig:Secuencia-Pulsos-K}, Qubit-Sensor Initialization).
We then correlate the qubit and the environment in the same way as
done when measuring the SENSIT contrast: we apply a $\left(\nicefrac{\pi}{2}\right)_{y}$
pulse to put the qubit-probe state in plane, then apply $L_{0}$ cycles
of the MREV-8 pulse sequence to create a state where the qubit-probe
and environment are correlated (see Fig. \ref{fig:Secuencia-Pulsos-K},
Sensor-Environment Correlation $H_{SE}$). This creates the correlated
state described in Eq. (\ref{eq:estado_correlacionado_inplane}).
We then apply a rotation pulse $\mathrm{e}^{i\phi I_{x}}$ on the
environment spins by an angle $\phi$ around the $x$ direction. To
then create a time-reversal echo, we apply a $\pi$-pulse on the qubit-probe
to effectively change the sign of the system-environment Hamiltonian
that creates the correlations, and apply again the MREV-8 sequence
during a time $T_{p}$ (see Fig. \ref{fig:Secuencia-Pulsos-K}, Sensor-Environment
Backwards Propagation $-H_{SE}$). This last step effectively creates
a backward evolution to refocus the initial state. After that, a time
reversal echo is created whose amplitude depends on $\phi$ to allow
encoding the number of correlated spins $K$ \citep{Alvarez2015,Dominguez2020,Dominguez2021,Niknam2020}.
Finally, the amplitude of the echo is measured, as shown in Fig. \ref{fig:Secuencia-Pulsos-K},
Qubit-Sensor Readout.

The total evolution operator of this time reversal quantum evolution
is thus $U_{\text{echo}}=\mathrm{e}^{iS_{z}BT_{p}}\mathrm{e}^{i\phi I_{x}}\mathrm{e}^{-iS_{z}BT_{p}}$.
The density matrix of Eq. (\ref{eq:estado_correlacionado_inplane})
can be written as $\rho(T_{p})=\sum_{M}\rho_{M}$, where $\rho_{M}$
are the multiple.quantum coherence blocks of order $M$, defined by
the property $\mathrm{e}^{i\phi I_{x}}\rho_{M}\mathrm{e}^{-i\phi I_{x}}=\mathrm{e}^{i\phi M}\rho_{M}$.
The effect of rotating the environmental spins around the $x$ axis
thus adds a different phase to each coherence component $\rho_{M}$
of the density matrix, i.e. $\mathrm{e}^{i\phi I_{x}}\rho(T_{p})\mathrm{e}^{-i\phi I_{x}}=\sum_{M}\mathrm{e}^{i\phi M}\rho_{M}$.
The measured echo is therefore 
\[
\left\langle \rho_{0}U_{\text{echo}}\rho_{0}U_{\text{echo}}^{\dagger}\right\rangle =\left\langle \rho(T_{p})\mathrm{e}^{i\phi I_{x}}\rho(T_{p})\mathrm{e}^{-i\phi I_{x}}\right\rangle =\sum_{M}\mathrm{e}^{i\phi M}\left\langle \rho_{M}^{\dagger}\rho_{M}\right\rangle \,.
\]
The dependence of this measured echo with respect to $\phi$ encodes
the MQC spectrum $\left\langle \rho_{M}^{\dagger}\rho_{M}\right\rangle $,
as they are multiplied by $\mathrm{e}^{i\phi M}$. The Fourier transform
of the echo signal with respect to the phase $\phi$ is the MQC spectrum
$\left\langle \rho_{M}^{\dagger}\rho_{M}\right\rangle $.  The effective
number of correlated spins $K$ is determined from the width of the
MQC spectrum, i.e. its second moment $K=\frac{\sum_{M}M^{2}\left\langle \rho_{M}^{\dagger}\rho_{M}\right\rangle ^{2}}{\sum_{M}\left\langle \rho_{M}^{\dagger}\rho_{M}\right\rangle ^{2}}=\left\langle \left[\rho,I_{x}\right]^{\dagger}\left[\rho,I_{x}\right]\right\rangle $,
where $\sum_{M}\left\langle \rho_{M}^{\dagger}\rho_{M}\right\rangle =\text{tr}\rho^{2}$
is constant \citep{Alvarez2015,Dominguez2020,Dominguez2021,Niknam2020}.
The second moment $K$ quantifies the norm of the commutator between
the localized initial state before the preparation time with the evolved
density matrix after the preparation time \citep{Dominguez2020,Dominguez2021}.
This thus determine a measure of distance to equilibrium based on
the number of correlated spins. Notice that this state does not commute
with the equilibrium state, thus relaxes to equilibrium during the
scrambling time. 

\end{widetext}

\end{document}